\documentclass[conference,a4paper]{IEEEtran}
\usepackage{xcolor}
\usepackage{balance}

\newcommand{\exportFigures}{true}
\usepackage{xcolor}

\usepackage{cite}
\usepackage{amsmath,amssymb,amsfonts,amsthm}
\usepackage{empheq}
\usepackage{algorithmic}

\usepackage{graphicx}

\usepackage[indexonlyfirst]{glossaries}
\makenoidxglossaries%
\newacronym{aoa}{AOA}{angle-of-arrival}
\newacronym{aod}{AOD}{angle-of-departure}
\newacronym{asic}{ASIC}{application specific integrated circuit}
\newacronym{awgn}{AWGN}{additive white Gaussian noise}
\newacronym{bs}{BS}{base station}
\newacronym{ccs}{CCS}{correlative channel sounder}
\newacronym{ce}{CE}{channel estimation}
\newacronym{csi}{CSI}{channel state information}
\newacronym{csp}{CSP}{contact service point}
\newacronym{crlb}{CRLB}{Cram\'er-Rao lower bound}
\newacronym{da}{DA}{data association}
\newacronym{dc}{DC}{direct current}
\newacronym{dm}{DM}{diffuse multipath}
\newacronym{dmc}{DMC}{\gls{dm} component}
\newacronym{ecsp}{ECSP}{Edge Computing Service Point}
\newacronym{eirp}{EIRP}{equivalent isotropically radiated power}
\newacronym{en}{EN}{energy neutral}
\newacronym{end}{END}{energy neutral device}
\newacronym{epu}{EPU}{edge processing unit}
\newacronym{fa}{FA}{federation anchor}
\newacronym{gbsc}{GBSC}{geometry-based deterministic stochastic channel}
\newacronym{gsm}{GSM}{Global System for Mobile Communications}  %
\newacronym{ic}{IC}{integrated circuit}
\newacronym{id}{ID}{information decoding}
\newacronym{iot}{IoT}{Internet of Things}
\newacronym{lis}{LIS}{large intelligent surface}
\newacronym{los}{LoS}{line-of-sight}
\newacronym{lsa}{LSA}{large synthetic array}
\newacronym{lti}{LTI}{linear time-invariant}
\newacronym{mmwave}{mmWave}{millimeter wave}
\newacronym{mimo}{MIMO}{multiple-input multiple-output}
\newacronym{miso}{MISO}{multiple-input single-output}
\newacronym{mpc}{MPC}{multipath component}
\newacronym{mrt}{MRT}{maximum ratio transmission}
\newacronym{nfc}{NFC}{near-field communication}
\newacronym{ofdm}{OFDM}{orthogonal frequency-division multiplexing}
\newacronym{ospa}{OSPA}{optimal subpattern assignment}
\newacronym{ota}{OTA}{over-the-air}
\newacronym{nb}{NB}{narrowband}
\newacronym{nlos}{NLoS}{non-line-of-sight}
\newacronym{pa}{PA}{power amplifier}
\newacronym{pl}{PL}{path loss}
\newacronym{pla}{PLA}{physically large array}
\newacronym{pc}{PC}{pilot count}
\newacronym{rcs}{RCS}{radar cross section}
\newacronym{rf}{RF}{radio frequency}
\newacronym{rfid}{RFID}{radio frequency identification}
\newacronym{ris}{RIS}{reflective intelligent surface}
\newacronym{rms}{RMS}{root mean square}
\newacronym{rss}{RSS}{received signal strength}
\newacronym{rssi}{RSSI}{received signal strength indicator}
\newacronym{ra}{RA}{reflectarray}
\newacronym{rw}{RW}{RadioWeaves}
\newacronym{sa}{SA}{synchronization anchor}
\newacronym{sar}{SAR}{specific absorption rate}
\newacronym{sbl}{SBL}{sparse Bayesian learning}
\newacronym{sdn}{SDN}{software-defined network}
\newacronym{sinr}{SINR}{signal-to-interference-plus-noise ratio}
\newacronym{simo}{SIMO}{single-input multiple-output}
\newacronym{siso}{SISO}{single-input single-output}
\newacronym{slam}{SLAM}{simultaneous localization and mapping}
\newacronym{snr}{SNR}{signal-to-noise ratio}
\newacronym{smc}{SMC}{specular multipath component}
\newacronym{spla}{SPLA}{synthetic \gls{pla}}
\newacronym{s-parameters}{S-parameters}{scattering parameters}
\newacronym{tosm}{TOSM}{through-open-short-match}
\newacronym{tdoa}{TDOA}{time-difference-of-arrival}
\newacronym{toa}{TOA}{time-of-arrival}
\newacronym{ue}{UE}{user equipment}
\newacronym{uhf}{UHF}{ultra high frequency}
\newacronym{ula}{ULA}{uniform linear array}
\newacronym{upa}{UPA}{uniform planar array}
\newacronym{ura}{URA}{uniform rectangular array}
\newacronym{uwb}{UWB}{ultrawideband}
\newacronym{vna}{VNA}{vector network analyzer}
\newacronym{wb}{WB}{wideband}
\newacronym{wpt}{WPT}{wireless power transfer}
\newacronym{zf}{ZF}{zero forcing}

\newacronym{cdf}{CDF}{cumulative distribution function}
\newacronym{ssf}{SSF}{small-scale fading}
\newacronym{lsf}{LSF}{large-scale fading}
\newacronym{ple}{PLE}{pathloss exponent}
\newacronym{usrp}{USRP}{universal software radio peripheral}
\newacronym{rb}{Rb}{Rubidium}
\newacronym{pps}{PPS}{pulse-per-second}
\newacronym{mrc}{MRC}{maximum ratio combinig}
\newacronym{fpga}{FPGA}{field-programmable gate array}
\newacronym{agc}{AGC}{automatic gain control}
\newacronym{tdma}{TDMA}{time-division multiple access}
\newacronym{cp}{CP}{cyclic prefix}
\newacronym{dac}{DAC}{digital-to-analog converter}

\newacronym{pdf}{PDF}{probability density function}
\newacronym{cbsm}{CBSM}{correlation-based stochastic model}
\newacronym{gbsm}{GBSM}{geometry-based stochastic model}
\newacronym{vr}{VR}{visibility region}
\newacronym{acf}{ACF}{autocorrelation function}

\usepackage{textcomp}

\usepackage{url}
\usepackage{bm}

\usepackage[caption=false,font=footnotesize]{subfig}

\usepackage[normalem]{ulem}

\usepackage{mathtools, cuted}

\usepackage{pdfpages} 

\usepackage{listings}
\usepackage[binary-units=true]{siunitx}

\usepackage{tikz}
\usetikzlibrary{angles}
\usetikzlibrary{arrows}
\usetikzlibrary{arrows.meta}
\usetikzlibrary{backgrounds}
\usetikzlibrary{calc}
\usetikzlibrary{decorations.pathmorphing}
\usetikzlibrary{fit}
\usetikzlibrary{patterns}
\usetikzlibrary{petri}
\usetikzlibrary{plotmarks}
\usetikzlibrary{positioning}
\usetikzlibrary{quotes}
\usetikzlibrary{shapes.misc}
\usetikzlibrary{spy}

\usepackage{pgfplots}
\pgfplotsset{compat=newest}
\usepgfplotslibrary{patchplots}
\usepgfplotslibrary{units}
\usepgfplotslibrary{polar}

\usepackage[most,theorems]{tcolorbox}
\tcbuselibrary{breakable}
\tcbset{every box/.style={enhanced,breakable}}

\usepackage[ruled,vlined]{algorithm2e}
\SetKwRepeat{Do}{do}{while}
\usepackage{setspace} 

\usepackage{cancel}

\usepackage{ifthen}

\newcommand{\tikzpng}[2]{
\ifthenelse{#1=1}
{\centering\input{#2.tex}}
{\centering\includegraphics[width=\figurewidth]{#2.png}}
}

\ifthenelse{\equal{\exportFigures}{true}}
{
  \usepgfplotslibrary{external}\tikzexternalize[prefix=tikz/]
}
{}

\newcommand{\ticked}{$\text{\rlap{$\checkmark$}}\square$}
\newcommand{\unticked}{{$\square$}}
\newcommand{\tick}[1]{\ifthenelse{#1=1}{\ticked}{\unticked}}

\newcounter{assump}

\definecolor{tw}{RGB}{51,183,150}

\newcommand{\rmv}{\hspace*{-.3mm}}

\newcommand{\vm}[1]{\ensuremath{\bm{#1}}} 

\newlength{\figurewidth}
\newlength{\figureheight}
\newlength{\belowFigureMargin}
\usepackage[a4paper,left=1.43cm,right=1.43cm,top=1.9cm,bottom=4.43cm]{geometry}

\ifthenelse{\equal{\exportFigures}{true}}
{
    \usepackage{tikz}
    \usetikzlibrary{external}
    \tikzexternalize[prefix=tikz/]
}
{}

\begin{document}
  
\title{{Propagation Modeling for Physically Large Arrays: \\ Measurements and Multipath Component Visibility}}

\author{\IEEEauthorblockN{
    Thomas Wilding\IEEEauthorrefmark{4},   
    Benjamin J. B. Deutschmann\IEEEauthorrefmark{4}, 
    Christian Nelson\IEEEauthorrefmark{3},
    Xuhong Li\IEEEauthorrefmark{3},\\
    Fredrik Tufvesson\IEEEauthorrefmark{3},
    Klaus Witrisal\IEEEauthorrefmark{4}
  }                                     
  \thanks{ The project has received funding from the European Union’s Horizon 2020 research and innovation programme under grant agreement No 101013425.}
  \IEEEauthorblockA{\IEEEauthorrefmark{4}
    Graz University of Technology, Austria, 
  \IEEEauthorrefmark{3}
    Lund University, Sweden,}
  \IEEEauthorblockA{ Email: \emph{\{thomas.wilding,benjamin.deutschmann,witrisal\}@tugraz.at} }
}

\maketitle

\begin{abstract}
This paper deals with propagation and channel modeling for physically large arrays. The focus lies on acquiring a spatially consistent model, which is essential, especially for positioning and sensing applications. Ultra-wideband, synthetic array measurement data have been acquired with large positioning devices to support this research. We present a modified multipath channel model that accounts for a varying visibility of multipath components along a large array. Based on a geometric model of the measurement environment, we analyze the visibility of specular components. We show that, depending on the size of the reflecting surface, geometric visibility and amplitude estimates obtained with a super-resolution channel estimation algorithm show a strong correspondence. Furthermore, we highlight the capabilities of the developed synthetic array measurement system.
\end{abstract}


\vskip0.5\baselineskip
\begin{IEEEkeywords}
  Propagation modeling, radio measurements, radio positioning.
\end{IEEEkeywords}

%

%

\glsresetall
\section{Introduction}\label{sec:introduction}

In recent years, the number of mobile devices, commonly termed \gls{ue}, and the size and number of infrastructure have increased. 
Such distributed infrastructure is very promising for future communication systems and has been widely researched in, e.g., XL-MIMO \cite{Guerra_Access2022}, radio stripes \cite{ShaikTC2021}, \gls{lis} \cite{DardariJSAC2020} or \gls{ris} \cite{BjoernsonSPM2022}, as well as \gls{rw} \cite{VanderPerreAsilomar2019,D1_1}.
In the context of \gls{rw}, the infrastructure is seen as federations of \glspl{csp}, highlighting its extended capabilities. The very large overall system aperture is envisioned to yield the required high system performance in terms of throughput, positioning accuracy, or wireless power transfer as well as efficiency.

Due to this large aperture, the channels for different regions of the \gls{csp} will become nonstationary.
In \cite{GaoAsilomar2013}, this nonstationarity was analyzed utilizing scattering cluster visibility regions for massive MIMO channels, similar to \cite{CarvalhoWC2020}, where types of nonstationarities and strategies to exploit these were discussed.
In \cite{FlordelisTWC2020} the COST2100 model was extended by visibility regions for scatterers, including a point process model for the birth and death of clusters.
When the employed channel model is not accurate enough, model-based estimators will suffer from errors due to model mismatch, which was analyzed in \cite{ChenArxiv2022}.

In this paper, we apply the notion of visibility regions to \glspl{mpc} and present an \gls{uwb} synthetic array measurement system that allows to collect data for propagation analysis and algorithm development.
We briefly outline a channel model including \gls{mpc} visibility and analyze the feasibility of this approach, performing non-parametric and parametric analyses of the measurements.

The remainder of this paper is structured as follows. 
Section~\ref{sec:system} outlines the channel model, Section~\ref{sec:measurements} introduces the developed measurement system, Section~\ref{sec:results} discusses the measurement results and relations to the channel model. Section~\ref{sec:conclusion} concludes the paper.

\section{System Model}\label{sec:system}

We consider a system with a single \gls{pla} characterized by a large aperture relative to the propagation distances of interest.
This is commonly referred to as propagation in the array near-field, where the wavefront curvature is noticeable, requiring accurate modeling for position related algorithms. 
\begin{figure}[t!]
  \centering
  \setlength{\figurewidth}{0.95\columnwidth}
  \setlength{\figureheight}{0.7\columnwidth}
  \input{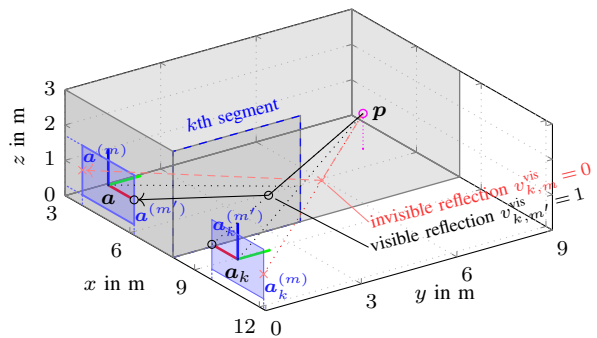}\vspace{-2mm}
  \caption{Visualization of a generic system setup, including a \gls{bs}-\gls{pla} at position $\vm{a}$, an exemplary mirror \gls{pla} $\vm{a}_k$, and a \gls{ue} at position $\vm{p}$. Exemplary multipath components corresponding to a wall segment of a limited extent are included, showing the resulting geometric visibility via ray-tracing.}
  \label{fig:setup}%
  \vspace{\belowFigureMargin}
\end{figure}
An illustration of the system is given in Fig.~\ref{fig:setup}, containing a single \gls{pla} representing a \gls{csp} at position $\bm{a}=[a_x,a_y,a_z]^\mathsf{T}$ that receives the signal from a single \gls{ue} at location $\bm{p}= [p_x,p_y,p_z]^\mathsf{T}$ equipped with a single antenna. 
The \gls{pla} is equipped with $M$ antenna elements at locations $\bm{a}^{(m)}=[a_x^{(m)},a_y^{(m)},a_z^{(m)}]^\mathsf{T}$, defined relative to the reference point $\vm{a}$.
The orientation of the \gls{pla} is known in the global coordinate system, with the orientation axes indicated as red, green, and blue lines.
To model deterministic, also termed specular, multipath propagation, we employ a mirror source model \cite{PedersenTAP2018} to model reflecting surfaces. 
An exemplary mirror source located at $\vm{a}_k$ with array elements at $\vm{a}_k^{(m)}$ (relative to $\vm{a}_k$) is included, obtained my mirroring the \gls{pla} at wall segment $k$ (dashed blue).
Note that the orientation at the mirror \gls{pla} is mirrored.
Considering a limited extent of the wall segment, not all array elements $m$ will receive a corresponding specular \gls{mpc}, indicated for array elements $m$ and $m'$ at the outer edges of the \gls{pla}.
The following section briefly outlines a signal model that accounts for the varying visibility.


\begin{figure*}[t!]\centering
  \subfloat[wall mounted mechanical positioner]{\centering
    \tikz{
      \node[inner sep=0mm] (fig) at (0,0) {\includegraphics[height=0.225\textwidth]{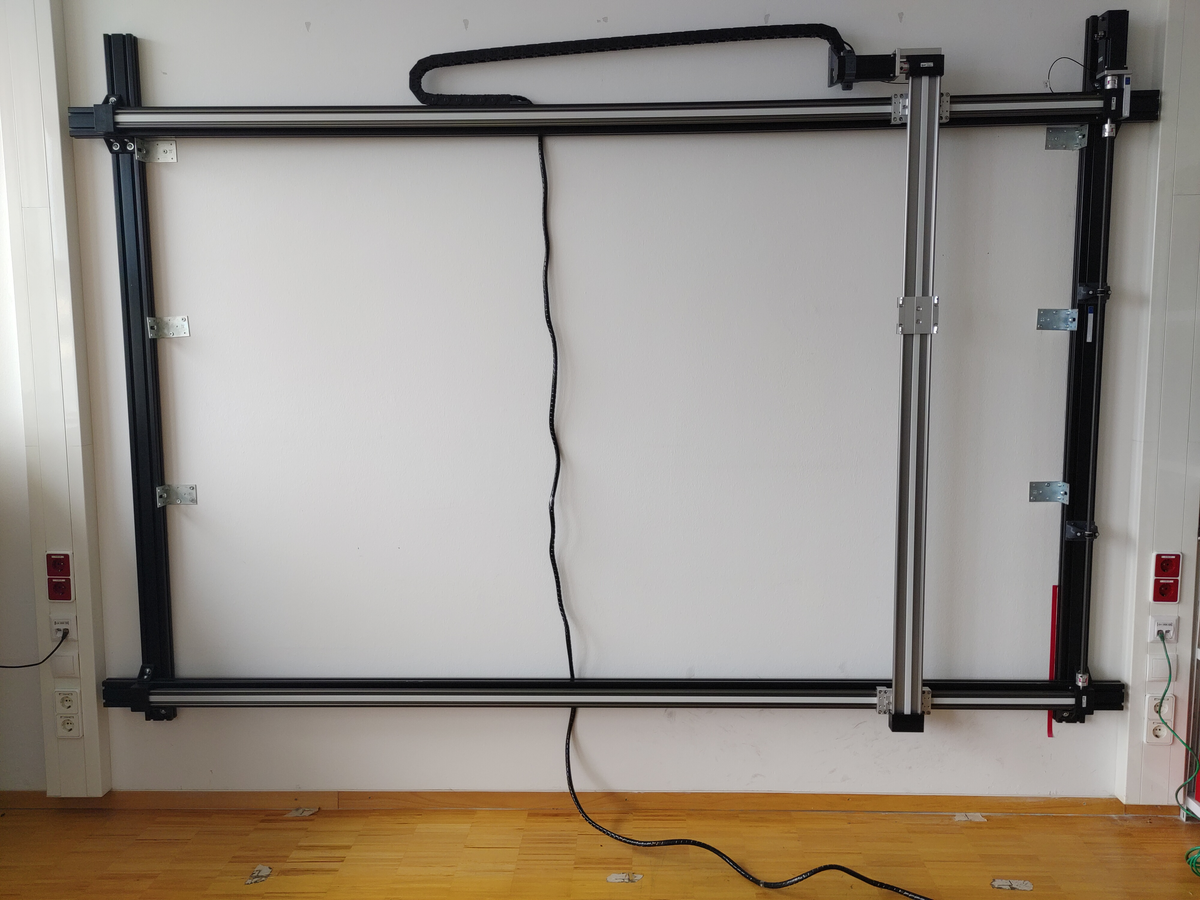}};
      \draw[red,<->,line width=0.75pt] ($(fig.north west)+(6.5mm,-6.5mm)$) -- ($(fig.north east)+(-5.5mm,-5.5mm)$) node[pos=0.5,anchor=north] {\footnotesize horizontal axis A};
      \draw[red,<->,line width=0.75pt] ($(fig.south west)+(7.5mm,10mm)$) -- ($(fig.south east)+(-6mm,10mm)$) node[pos=0.5,anchor=south,sloped] {\footnotesize horizontal axis B};
      \draw[blue,<->,line width=0.75pt] ($(fig.south east)+(-12mm,9mm)$) -- ($(fig.north east)+(-11mm,-3.5mm)$) node[pos=0.5,anchor=north,sloped] {\footnotesize vertical axis};
      \node[anchor=west] (s1) at (fig.center) {\footnotesize slide};
      \draw[black,->,line width=0.75pt] ($(s1.east)$) -- ($(s1.east)+(5mm,5mm)$);
    }
    \label{fig:env:wall}}
  \hspace{2mm}
  \subfloat[antenna and absorber]{\centering
    \includegraphics[height=0.225\textwidth]{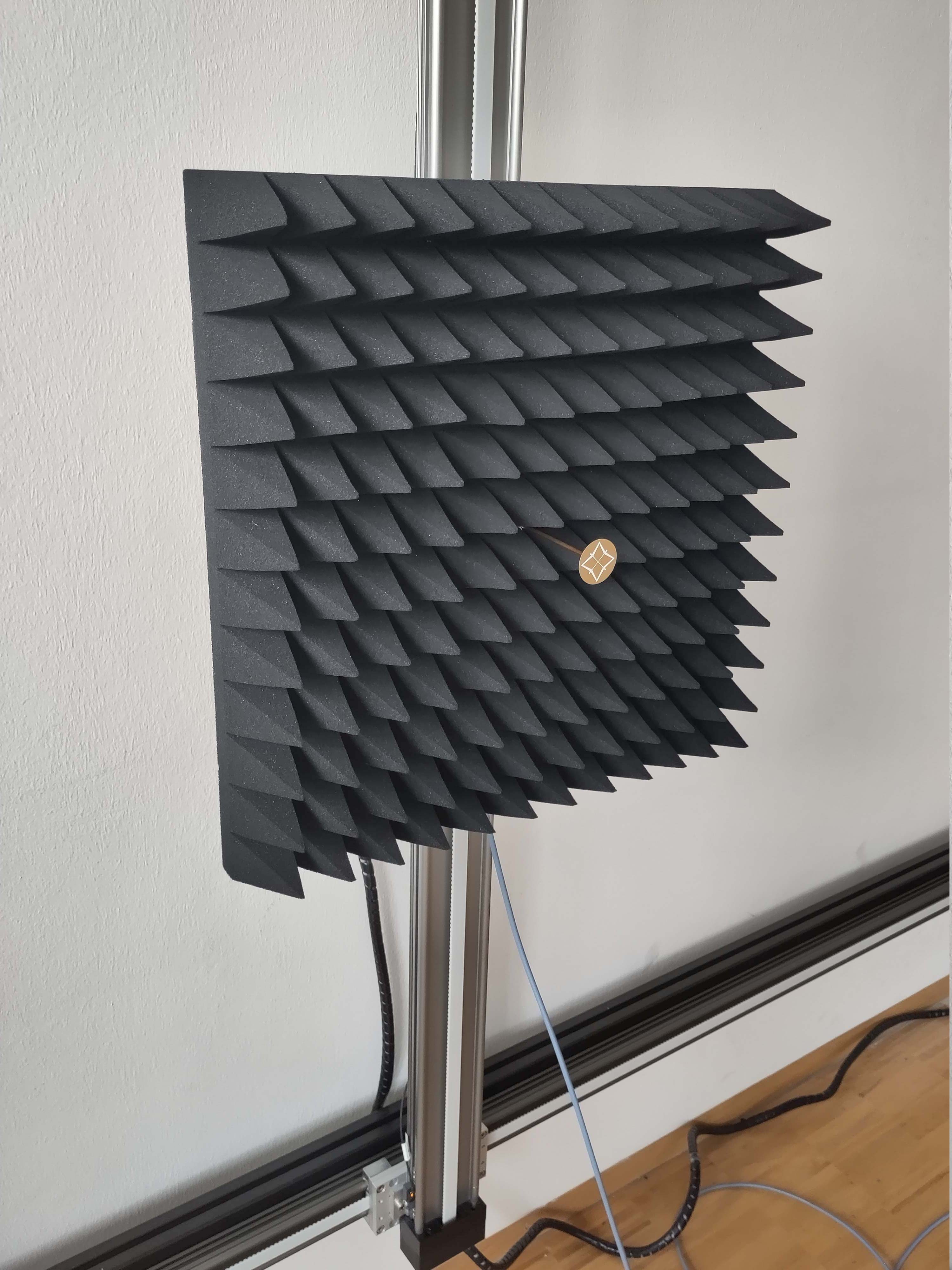}\label{fig:env:absorber}}
  \hspace{2mm}
  \subfloat[medium sized environment (lab/office room)]{\centering\hspace{-7mm}
   \setlength{\figurewidth}{0.45\textwidth}
   \setlength{\figureheight}{0.5\textwidth}
   \def\datapath{./figures/environment/demoroom_static}
   \input{\datapath/demoroom_static.tex}
   \label{fig:env:demoroom}}

  \subfloat[\gls{pla}, photo A in~\protect\subref{fig:env:corridor}]{
    \tikz{\node[inner sep=0pt] (fig) at (0,0) {\includegraphics[height=0.425\columnwidth,trim=150 0 150 800,clip]{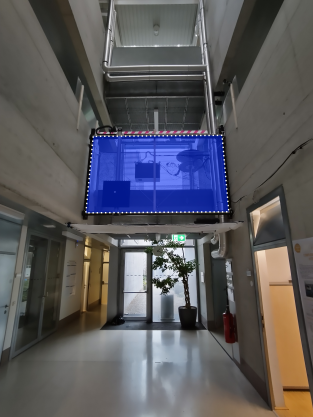}};
    \node[white,align=center,font=\scriptsize] at (0,10mm) {measurement area \\ $(2.26\,\mathrm{m}\times 1.55\,\mathrm{m})$};
    \draw[line width=0.75pt,white,<->] (-5mm,6mm) -- (-5mm,-15mm) node[anchor=west,font=\scriptsize] {$h=2.85\,\mathrm{m}$};
    \node[anchor=north east, inner sep=0.75pt, rotate=90, font=\scriptsize, white] at (fig.north west) {right wall};
    \node[anchor=south east, inner sep=0.75pt, rotate=90, font=\scriptsize, white] at (fig.north east) {left wall};
    }
    \label{fig:env:to2}}
  %
  %
  \subfloat[\gls{ue} region, photo B in~\protect\subref{fig:env:corridor}]{
    \centering
    \tikz{\node[inner sep=0pt] (fig) at (0,0) {\includegraphics[height=0.425\columnwidth,trim=480 100 0 200,clip]{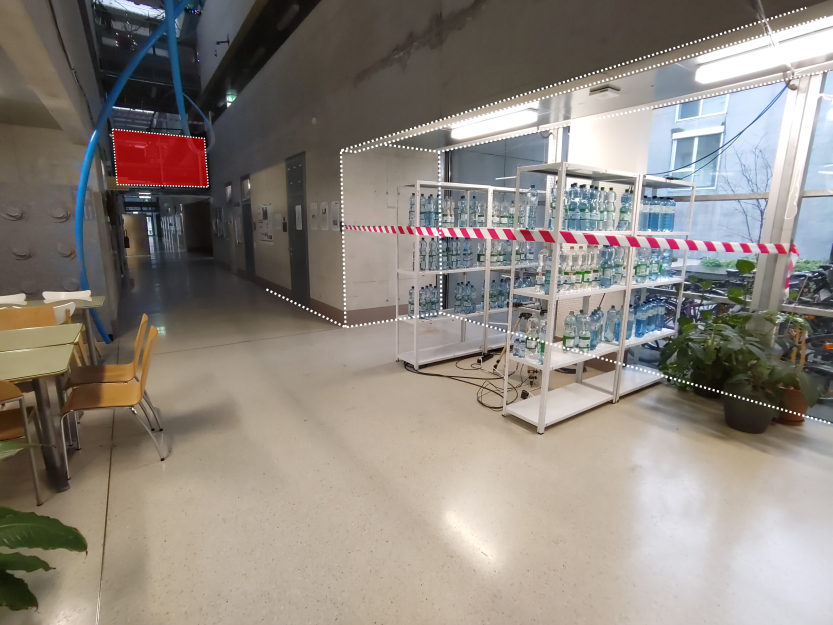}};
    \node[align=center,font=\scriptsize,black,fill=none,opacity=0.5,text opacity=1,rotate=-25] at ($(fig.center)+(0,-7mm)$) {shelves};
    \node[anchor=north east, inner sep=0.75pt, rotate=15, font=\scriptsize, white] at (fig.north) {right wall};
    }
    \label{fig:env:eg}}
  %
  %
  \subfloat[large sized environment (corridor)]{\hspace{-7.5mm}
    \setlength{\figurewidth}{1.2\columnwidth}
    \setlength{\figureheight}{1.2\columnwidth}
    \def\datapath{./figures/environment/corridor_locations}
    \input{\datapath/corridor_locations.tex}
    \label{fig:env:corridor}}

  \caption{Visualization of measurement locations in medium and large size environments. (a) points distributed in medium size environment, (b) trajectory including obstruction in medium size environment, (c) distributed points and trajectory in large size environment.}  
  \label{fig:env}%
  \vspace{\belowFigureMargin}
\end{figure*} 

The signal model includes specular \glspl{mpc} as well as scattering and diffuse propagation in the environment. 
At each frequency, the received signal vector $\bm{y}(f)$ for a baseband frequency $f$ is denoted by
\begin{align}
\vm{y}(f) &= \sum_{k=1}^K \vm{h}_k(f) s(f) +  \vm{w}_\text{s}(f) + \vm{w}(f) \in \mathbb{C}^{M}
\label{eq:signal_model}
\end{align}
consisting of a deterministic signal component in the form of a (finite) sum of $K$ \glspl{mpc} attributable to specific environment features, a stochastic signal component $\vm{w}_\text{s}$ and \gls{awgn} $\vm{w}(f)$, representing the . 
The latter represents the sampled receiver noise, plus the a stochastic signal-related component denoted by $\vm{w}_\text{s}$.
The stochastic signal-related component in $\vm{w}_\text{s}$ is commonly used to represent any form of stochastic multipath propagation such as scattering \cite{SchubertTAP2013,FlordelisTWC2020} or a dense multipath component \cite{RichterPhD2005,SalmiTSP2009,LeitingerJSAC2015}.

In the deterministic signal component, each \gls{mpc} $k$ is described by the channel vector  $\vm{h}_k(f) \in \mathbb{C}^{M}$ and the known transmit signal ${s}(f)$ in complex baseband. 
In realistic environments, the number of \glspl{mpc} $K$ is unknown and varies depending on the locations of the transmitting and receiving devices.
The entries of the channel vector are defined as \cite{johnson1992}
\begin{align}
  [\vm{h}_k(f)]_{m} = {\alpha}_{k,m}\mathrm{exp}(-j2\pi (f+f_\mathrm{c})\tau_{k,m})\label{eq:channel_model}
\end{align} 
with carrier frequency $f_\mathrm{c}$, \gls{mpc} delay $\tau_{k,m}=\| \vm{a}^{(m)}_k-\vm{p}\|/c$, speed of light $c$, and the \gls{mpc} amplitudes ${\alpha}_{k,m}$ defined as 
\begin{align}
  {\alpha}_{k,m} &\propto \frac{\tilde{\alpha}_{k,m}b(\varphi_{k,m},\theta_{k,m})}{c\tau_{k,m}}v^\text{vis}_{k,m}.\label{eq:amplitude_model}
\end{align}
The factor $b(\varphi,\theta)\in \mathbb{C}$ represents the complex-valued antenna gain pattern in direction of the azimuth $\varphi$ and elevation $\theta$ of the corresponding \gls{mpc} at the $m$th array element. 
The quantity $\tilde{\alpha}_{k,m}$ on the right-hand side of \eqref{eq:amplitude_model} represents the environment-related, path loss compensated \gls{mpc} amplitude, i.e., the amplitude related to transmit power and attenuation due to reflection at different materials.
Note that for a specific \gls{mpc} $k$, the amplitude $\tilde{\alpha}_{k,m}$ can vary per array element, e.g., due to angle dependent reflection coefficients of surfaces and their electromagnetic material properties, or when the size of the array becomes large, such that classical array processing assumptions such as plane wave propagation and negligible propagation attenuation along the array do not apply to the full \gls{pla}.
The visibility is considered by the factor $v^\text{vis}_{k,m}$, which is $v^\text{vis}_{k,m}=1$ if the component $k$ is visible at array element $m$, and $v^\text{vis}_{k,m}=0$ otherwise. 
Note that this factor can be absorbed into the amplitudes but is expressed here to highlight the relation to the environment.

\section{Measurement System}\label{sec:measurements}

The developed measurement system for synthetic \glspl{pla} consists of a mechanical positioner allowing to form 2-dimensional arrays with arbitrary planar geometry and a Rohde \& Schwarz ZVA24 \gls{vna} equipped with \gls{uwb} antennas for performing the channel measurements.

\paragraph{RF hardware and antennas}
The \gls{vna} covers a frequency range of \SI{24}{\giga\hertz} (from DC) and was calibrated with a \gls{tosm} calibration kit. The antennas used in the measurements are dipole antennas, manufactured from cent coins (see \cite[App.\,B.3]{KrallPhD2008}), and dipole-slot antennas that were manufactured according to the XETS-antenna design from \cite{CostaTAP2009}.
While the \gls{vna} covers a wider frequency range, measurements are only performed in the frequency band of $3-10\,\mathrm{GHz}$ for which the antennas are designed.
Due to the measurement aperture allowing reception from azimuth and elevation, the slot antennas' design was chosen as it exhibits an approximately omnidirectional pattern.
When mounted on the absorber and reflector plate construction, the antennas only receive from the half-space facing away from the absorber.

\paragraph{Mechanical positioner}
The mechanical positioner allows to form arbitrary planar synthetic array geometries in an automated fashion with sub-millimeter accuracy for the relative antenna positions.
The maximum measurement area that can be used spans roughly\footnote{Depending on the installation in the environment, the actual usable measurement area can be smaller due to necessary safety margins.} $2.5\,\mathrm{m} \times 1.5\,\mathrm{m}$, shown in Fig.~\ref{fig:env:wall}.
The horizontal and vertical axes are equipped with a \textit{dryve\textsuperscript{\textregistered}\,D1} motor controller driving two \textit{drylin\textsuperscript{\textregistered}\,NEMA24} stepper motors, allowing motion with variable acceleration, deceleration, and velocity.
Each stepper motor has a holding break to keep the position during stop intervals.
The vertical axis is equipped with a slide to mount the antennas and is itself attached to the horizontal axes.
To emulate a \gls{pla} mounted directly on a surface, e.g., a \gls{csp} in a \gls{rw} system, the mechanical positioner is equipped with a wideband absorber and a reflector plate on the slide behind the antenna (see Fig.~\ref{fig:env:absorber}), removing the reflection from the surface on which the positioner is mounted on.
Due to this construction, the distance from the antenna phase center to the back wall is roughly \SI{43}{\centi\metre} (including the antenna length of \SI{10}{\centi\metre}). 

\paragraph{Measurement environments}
Measurements were performed in two environments of different sizes: in a medium size room representing a typical lab/office environment and in a large environment with a high ceiling and larger surfaces.
Photos of measurement regions of the \glspl{pla} are shown in Fig.~\ref{fig:env:wall} and \ref{fig:env:to2}.
The number of array elements per \gls{pla} were $(112 \times 75)$ and $(88 \times 32)$, forming \glspl{ura} with $\lambda/2$-spacing at $f_\mathrm{c}=\SI{6.95}{\giga\hertz}$ and $f_\mathrm{c}=\SI{6}{\giga\hertz}$ in the medium and large environments, respectively, with $\lambda=c/f_\text{c}$ denoting the wavelength. 
In each environment, measurements at five \gls{ue} positions have been conducted, which are labeled Mx and Lx with antenna heights of $h_\text{M}=\{1.546, 0.895, 2.235, 1.478, 1.202\}\,\mathrm{m}$ and $h_\text{L}=\{1.145, 1.317, 1.162, 1.590, 1.592\}\,\mathrm{m}$, respectively. 
To aid the measurement analysis, 3-dimensional models were generated for both environments, shown in Fig.~\ref{fig:env:demoroom} and \ref{fig:env:corridor} for medium and large size, respectively.  
These allow to make use of the mirror source model to represent the \glspl{mpc} from \eqref{eq:signal_model}. 
The use of a mirror source model is widespread in literature \cite{PedersenTAP2018}, and in the context of this work enables to compare the model-based component visibility, with estimated components obtained by position-based beamforming or a super-resolution channel estimation algorithm.

\section{Measurement Results and Analysis}\label{sec:results}



\begin{table*}[t]
  \caption{Channel estimation results for all measurement positions in both measurement environments (see Fig.~\ref{fig:env}), using subarrays of size $(4 \times 4)$ and a bandwidth of \SI{500}{\mega\hertz}.}
  \label{tab:components-sbl}
\scriptsize
\begin{tabular}{|l|r|r|r|r|r||r|r|r|r|r|r|}\hline
  & \multicolumn{10}{c|}{component energy in \% of received signal energy (as mean $\pm$ std.dev.)} \\\hline
  component          & M1              & M2              & M3              & M4              & M5              & L1               & L2              & L3               & L4               & L5              \\\hline\hline
  strongest          & 85.8 $\pm$  6.5 & 76.5 $\pm$ 15.3 & 67.2 $\pm$  9.4 & 43.3 $\pm$ 14.6 & 85.4 $\pm$ 17.5 & 42.4 $\pm$ 11.5  & 26.4 $\pm$  9.0 & 12.7 $\pm$  7.0  & 25.6 $\pm$  6.5  &  9.3 $\pm$  3.8 \\
  2nd strong.      &  3.9 $\pm$  1.9 &  4.7 $\pm$  4.6 &  9.7 $\pm$  5.4 & 16.1 $\pm$  8.2 &  6.7 $\pm$  9.5 & 24.0 $\pm$  7.9  & 12.3 $\pm$  4.9 &  6.4 $\pm$  2.5  & 17.2 $\pm$  4.2  &  5.5 $\pm$  2.1 \\
  3rd                &  2.1 $\pm$  1.3 &  1.4 $\pm$  1.3 &  3.9 $\pm$  2.3 &  6.7 $\pm$  4.1 &  1.2 $\pm$  1.5 &  8.5 $\pm$  3.8  &  7.7 $\pm$  3.0 &  4.3 $\pm$  1.6  & 11.8 $\pm$  3.5  &  3.6 $\pm$  1.2 \\
  4th                &  1.1 $\pm$  0.6 &  0.7 $\pm$  0.7 &  2.4 $\pm$  1.0 &  4.2 $\pm$  1.9 &  0.6 $\pm$  0.6 &  4.5 $\pm$  2.4  &  4.8 $\pm$  1.7 &  3.3 $\pm$  1.3  &  7.2 $\pm$  3.6  &  2.6 $\pm$  0.8 \\
  5th                &  0.8 $\pm$  0.3 &  0.5 $\pm$  0.4 &  1.8 $\pm$  0.7 &  3.1 $\pm$  1.3 &  0.4 $\pm$  0.4 &  2.2 $\pm$  1.2  &  3.4 $\pm$  1.6 &  2.6 $\pm$  0.9  &  3.6 $\pm$  1.7  &  2.0 $\pm$  0.6 \\
  6th                &  0.6 $\pm$  0.3 &  0.4 $\pm$  0.3 &  1.3 $\pm$  0.5 &  2.4 $\pm$  0.9 &  0.3 $\pm$  0.3 &  1.3 $\pm$  0.7  &  2.3 $\pm$  1.0 &  2.0 $\pm$  0.7  &  2.3 $\pm$  1.1  &  1.7 $\pm$  0.5 \\\hline
  $\sum$ over 20     & \textbf{96.5} $\pm$  1.1 & 97.8 $\pm$  1.6 & \textbf{91.1} $\pm$  2.8 & 82.3 $\pm$  4.7 & \textbf{97.8} $\pm$  1.7 & 87.0 $\pm$  3.6  & 63.1 $\pm$  7.1 & 38.4 $\pm$  8.9  & 69.7 $\pm$  6.3  & 31.8 $\pm$  6.7 \\
  residual           &  3.1 $\pm$  1.0 &  1.9 $\pm$  1.4 &  7.9 $\pm$  2.6 & 15.9 $\pm$  4.2 &  1.9 $\pm$  1.5 & 12.3 $\pm$  3.4  & 35.3 $\pm$  6.8 & 60.9 $\pm$ 15.3  & 29.0 $\pm$  6.1  & 65.4 $\pm$  6.5 \\\hline
\end{tabular}

  %
  \vspace{\belowFigureMargin}
\end{table*}

This section presents the results from the performed measurement analysis and summarizes the results for the models in \eqref{eq:signal_model}, \eqref{eq:channel_model} and \eqref{eq:amplitude_model}.
For the processing of measurements, we compare non-parametric and parametric approaches, both aided by a mirror source model based on the environment models (see Fig.~\ref{fig:env}). 
The non-parametric approach is a standard spherical wave beamformer, computed for the entire array and a bandwidth of \SI{3}{\giga\hertz}. 
The parametric approach is a super-resolution \gls{sbl} algorithm to estimate \glspl{mpc}, based on \cite{HansenSAM2014}.
For processing the measurements with the \gls{sbl} algorithm, a bandwidth of \SI{500}{\mega\hertz} is used and the \gls{pla} is separated into subarrays of size ($4\times 4$) or ($8\times 8$).
This allows to make use of the plane-wave assumption per subarray and to analyse the \gls{mpc} visibility and amplitudes.

\subsection{Spherical wave beamforming}\label{sec:spherical_beamformer}
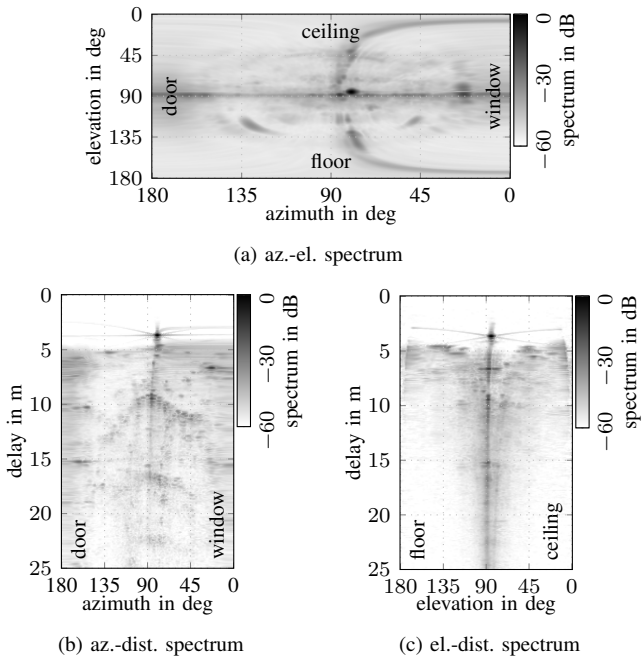
\begin{figure}[t!]
  \centering
  \subfloat[az.-el. spectrum]{
  \setlength{\figurewidth}{0.26\textwidth}
  \setlength{\figureheight}{0.12\textwidth}
  \def\datapath{./figures/results/beamformer/demoroom_pos1/sphbeamf_az_el_img}

%
%
 
\pgfplotsset{every axis/.append style={
  label style={font=\footnotesize},
  legend style={font=\footnotesize},
  tick label style={font=\footnotesize},
  xticklabel={
    \ifdim \tick pt < 0pt
      \pgfmathparse{abs(\tick)}%
      \llap{$-{}$}\pgfmathprintnumber{\pgfmathresult}
   \else
      \pgfmathprintnumber{\tick}
   \fi
}}}
 
\begin{tikzpicture}

\begin{axis}[%
width=\figurewidth,
height=\figureheight,
at={(0\figurewidth,0\figureheight)},
scale only axis,
point meta min=-60,
point meta max=0,
axis on top,
x dir=reverse,
xmin=-90.2506963788301,
xmax=90.2506963788301,
xlabel={azimuth in deg},
xtick={-90,-45,0,45,90},
xticklabels={0,45,90,135,180},
y dir=reverse,
ymin=-0.502793296089385,
ymax=180.502793296089,
ylabel={elevation in deg},
ytick={0,45,90,...,180},
axis background/.style={fill=white},
xmajorgrids,
ymajorgrids,
major tick length=0.75mm,
grid style=dotted,
legend style={legend cell align=left, align=left, draw=white!15!black},
colormap={mymap}{[1pt] rgb(0pt)=(1,1,1); rgb(255pt)=(0,0,0)},
ylabel style={yshift=-2mm},
xlabel style={yshift=2mm},
colorbar,
colorbar style={at={(1.01,1)},anchor=north west,major tick length=1mm,width=1.75mm,height=1.75cm,ylabel=spectrum in dB, yticklabel style={rotate=90,font=\footnotesize,xshift=-0.5mm}, ylabel style={yshift=1.5mm},ytick distance={30}},
]
\addplot [forget plot] graphics [xmin=-90.2506963788301, xmax=90.2506963788301, ymin=-0.502793296089385, ymax=180.502793296089] {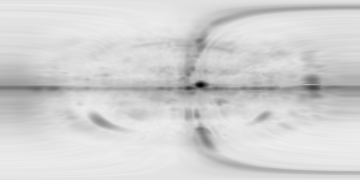};

\node[font=\footnotesize,anchor=south,rotate=90] at (axis cs: -90,90) {window};
\node[font=\footnotesize,anchor=north,rotate=90] at (axis cs:  90,90) {door};
\node[font=\footnotesize,anchor=north,rotate=0] at (axis cs:   0, 0) {ceiling};
\node[font=\footnotesize,anchor=south,rotate=0] at (axis cs:   0,180) {floor};

\end{axis}
\end{tikzpicture}%
  \label{fig:meas:tug:medium:pos1:sph:az_el}
  }\vspace{-2mm}

  \subfloat[az.-dist. spectrum]{\hspace{-4mm}
    \setlength{\figurewidth}{0.125\textwidth}
    \setlength{\figureheight}{0.2\textwidth}
    \def\datapath{./figures/results/beamformer/demoroom_pos1/sphbeamf_az_d_img}

%
%
 
\pgfplotsset{every axis/.append style={
  label style={font=\footnotesize},
  legend style={font=\footnotesize},
  tick label style={font=\footnotesize},
  xticklabel={
    \ifdim \tick pt < 0pt
      \pgfmathparse{abs(\tick)}%
      \llap{$-{}$}\pgfmathprintnumber{\pgfmathresult}
   \else
      \pgfmathprintnumber{\tick}
   \fi
}}}
 
\begin{tikzpicture}

\begin{axis}[%
width=\figurewidth,
height=\figureheight,
at={(0\figurewidth,0\figureheight)},
scale only axis,
point meta min=-60,
point meta max=0,
axis on top,
x dir=reverse,
xmin=-90.2506963788301,
xmax=90.2506963788301,
xlabel={azimuth in deg},
xtick={-90,-45,0,45,90},
xticklabels={0,45,90,135,180},
ytick distance={5},
y dir=reverse,
ymin=-0.0468425715625,
ymax=25,
ylabel={delay in m},
axis background/.style={fill=white},
xmajorgrids,
ymajorgrids,
major tick length=0.75mm,
grid style=dotted,
legend style={legend cell align=left, align=left, draw=white!15!black},
colormap={mymap}{[1pt] rgb(0pt)=(1,1,1); rgb(255pt)=(0,0,0)},
ylabel style={yshift=-2mm},
xlabel style={yshift=2mm},
colorbar,
colorbar style={at={(1.02,1)},anchor=north west,major tick length=1mm,width=1.75mm,height=1.75cm,ylabel=spectrum in dB, yticklabel style={rotate=90,font=\footnotesize,xshift=-0.5mm}, ylabel style={yshift=1.5mm},ytick distance={30}},
]
\addplot [forget plot] graphics [xmin=-90.2506963788301, xmax=90.2506963788301, ymin=-0.0468425715625, ymax=39.9567135428125] {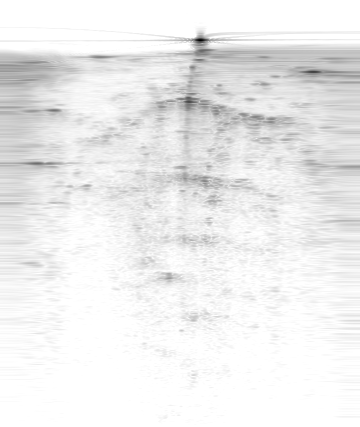};

\node[font=\footnotesize,anchor=south west,rotate=90] at (axis cs: -90,25) {window};
\node[font=\footnotesize,anchor=north west,rotate=90] at (axis cs:  90,25) {door};

\end{axis}
\end{tikzpicture}%
    \label{fig:meas:tug:medium:pos1:sph:az_d}
  }
  \subfloat[el.-dist. spectrum]{\hspace{-4mm}
    \setlength{\figurewidth}{0.125\textwidth}
    \setlength{\figureheight}{0.2\textwidth}
    \def\datapath{./figures/results/beamformer/demoroom_pos1/sphbeamf_el_d_img}

%
%
 
\pgfplotsset{every axis/.append style={
  label style={font=\footnotesize},
  legend style={font=\footnotesize},
  tick label style={font=\footnotesize},
  xticklabel={
    \ifdim \tick pt < 0pt
      \pgfmathparse{abs(\tick)}%
      \llap{$-{}$}\pgfmathprintnumber{\pgfmathresult}
   \else
      \pgfmathprintnumber{\tick}
   \fi
}}}
 
\begin{tikzpicture}

\begin{axis}[%
width=\figurewidth,
height=\figureheight,
at={(0\figurewidth,0\figureheight)},
scale only axis,
point meta min=-60,
point meta max=0,
axis on top,
x dir=reverse,
xmin=-0.502793296089385,
xmax=180.502793296089,
xtick={0,45,90,...,180},
xlabel={elevation in deg},
y dir=reverse,
ymin=-0.0468425715625,
ymax=25,
ytick distance={5},
ylabel={delay in m},
axis background/.style={fill=white},
xmajorgrids,
ymajorgrids,
major tick length=0.75mm,
grid style=dotted,
legend style={legend cell align=left, align=left, draw=white!15!black},
colormap={mymap}{[1pt] rgb(0pt)=(1,1,1); rgb(255pt)=(0,0,0)},
ylabel style={yshift=-2mm},
xlabel style={yshift=2mm},
colorbar,
colorbar style={at={(1.02,1)},anchor=north west,major tick length=1mm,width=1.75mm,height=1.75cm,ylabel=spectrum in dB, yticklabel style={rotate=90,font=\footnotesize,xshift=-0.5mm}, ylabel style={yshift=1.5mm},ytick distance={30}},
]
\addplot [forget plot] graphics [xmin=-0.502793296089385, xmax=180.502793296089, ymin=-0.0468425715625, ymax=39.9567135428125] {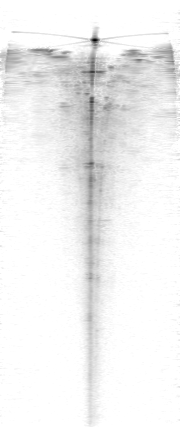};

\node[font=\footnotesize,anchor=south west,rotate=90] at (axis cs: 0,25) {ceiling};
\node[font=\footnotesize,anchor=north west,rotate=90] at (axis cs: 180,25) {floor};

\end{axis}
\end{tikzpicture}%
    \label{fig:meas:tug:medium:pos1:sph:el_d}
  }
  
  \caption{Spherical wave beamformer spectra for position M1 in the medium size environment using the full data (full bandwidth, all array elements). The azimuth-elevation power spectrum (Fig.~\ref{fig:meas:tug:medium:pos1:sph:az_el}) represents the view from the synthetic \gls{pla} into the room (see Fig.~\ref{fig:env:demoroom}). 
  }\label{fig:env:medium:pos1:sph}%
  \vspace{\belowFigureMargin}
\end{figure}

An exemplary spherical beamformer spectrum computed for position M1 is shown in Fig.~\ref{fig:env:medium:pos1:sph} in terms of the marginal spectra for combinations of delay, azimuth, and elevation.
The azimuth-elevation power spectrum in Fig.~\ref{fig:meas:tug:medium:pos1:sph:az_el} represents the view from the \gls{pla} shown in Fig.~\ref{fig:env:demoroom}. 
The azimuth-elevation spectrum in Fig.~\ref{fig:meas:tug:medium:pos1:sph:az_el} shows a number of specular \glspl{mpc} in addition to diffuse components.
The azimuth-delay and elevation-delay spectra show similar combinations of specular and diffuse components.
Locations of high power in the full spectrum represent locations of origin of specular \glspl{mpc} from~\eqref{eq:channel_model}, i.e., of the corresponding mirror sources.
A downside of the use of the full array for spherical beamforming is that the visibility information of \glspl{mpc} is lost, with partially visible \glspl{mpc} simply resulting in lower power at the corresponding spectrum bin. 

 \begin{figure*}[tp!]
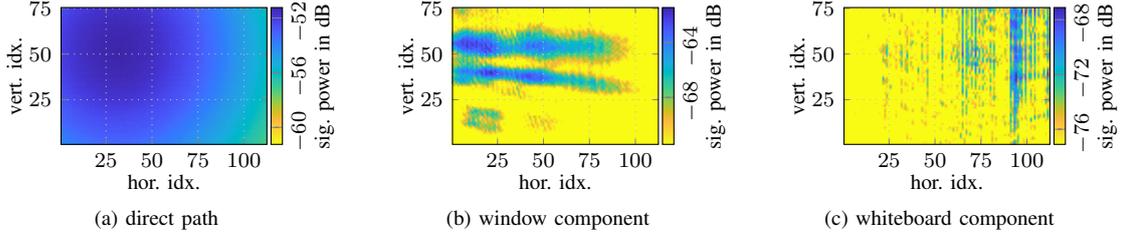

 \centering
  \subfloat[direct path]{
    \setlength{\figurewidth}{0.25\textwidth}
    \setlength{\figureheight}{0.1\textwidth}
    \def\datapath{./figures/results/model_based/pos1_k1_power_distribution_3000MHz_noPLcomp/}
    \input{\datapath/pos1_k1_power_distribution_3000MHz_noPLcomp.tex}
    \label{fig:geometric_analysis:pos1:direct}
  } 
  \subfloat[window component]{
    \setlength{\figurewidth}{0.25\textwidth}
    \setlength{\figureheight}{0.1\textwidth}
    \def\datapath{./figures/results/model_based/pos1_k5_power_distribution_3000MHz_noPLcomp/}
    \input{\datapath/pos1_k5_power_distribution_3000MHz_noPLcomp.tex}
    \label{fig:geometric_analysis:pos1:window}
  }
  \subfloat[whiteboard component]{
    \setlength{\figurewidth}{0.25\textwidth}
    \setlength{\figureheight}{0.1\textwidth}
    \def\datapath{./figures/results/model_based/pos1_k4_power_distribution_3000MHz_noPLcomp/}
    \input{\datapath/pos1_k4_power_distribution_3000MHz_noPLcomp.tex}
    \label{fig:geometric_analysis:pos1:whiteboard}
  }
  
  \caption{Signal power for each antenna position (e.g., \gls{pla} element) computed by performing position-based, i.e., spherical wave, beamforming for the  direct path, window, and whiteboard reflections at the example of position M1. The (image) source positions are computed from the model in Fig.~\ref{fig:env:demoroom}. 
  }\label{fig:geometric_analysis:pos1}%
  \vspace{\belowFigureMargin}
\end{figure*}

\subsection{Geometry-based analysis}\label{sec:geometry-based}
By exploiting the known measurement position in combination with the environment model (at the example of position M1), the received power per-array element is computed to show local variations over the area of the \gls{pla}.
Results are shown in Fig.~\ref{fig:geometric_analysis:pos1} for three components: the direct path, the window reflection, and the whiteboard reflection, with the corresponding environment segments highlighted by red, dashed outlines in Fig.~\ref{fig:env:demoroom}.
The direct path amplitudes in Fig.~\ref{fig:geometric_analysis:pos1:direct} show the expected attenuation due to the distance-dependent path loss and the gain pattern of the antenna. 
As patch antennas were used at the \gls{pla} and the \gls{ue}, the gain pattern affects the amplitudes twice. 
The full \gls{pla} is in \gls{los} condition with respect to the direct path.
The window component (see Fig.~\ref{fig:geometric_analysis:pos1:window}) shows much stronger amplitude variations and a distinct visibility region. 
The whiteboard component (see Fig.~\ref{fig:geometric_analysis:pos1:whiteboard}) shows significant amplitude fading, which relates to the computed geometric visibility regions (c.f. Fig.~\ref{fig:meas:tug:static:medium:pos1:VA5:vis}).

\subsection{Super-resolution channel estimation}
\label{sec:sbl}
For the parametric analysis, the \gls{sbl} algorithm \cite{HansenSAM2014} is applied to the signals per subarray.
This allows to make use of the plane wave assumption and assume negligible propagation attenuation along the subarrays \cite{johnson1992}, due to the small subarray size w.r.t. the propagation distance to modify \eqref{eq:signal_model} accordingly.
Furthermore, components are assumed to be visible for all subarray elements, using square \glspl{ura} with $M=\{16,64\}$ array elements.
Increasing $M$ improves the analysis of the component visibility due to the array gain, e.g., compared to the results from Sec.~\ref{sec:geometry-based}. 
We restrict the maximum number of components to be estimated in the \gls{sbl}-algorithm to $\hat{K}=20$ to reduce the computation time.

\subsubsection{Model representation}
A summary of the \gls{sbl}-estimates for the measurement positions in both environments is given in Table~\ref{tab:components-sbl}.
The table compares the estimated component energy (as a percentage of the total received signal energy) for the six strongest components at all measurement positions. 
The results are shown in terms of mean and standard deviation taken over the resulting $N_\text{s}=500$ subarrays.
Comparing the total estimated energy with the residual energy gives a metric on how well the channel estimator captures the total signal energy.
The results for different measurement positions show that a high percentage of the signal energy is covered by all estimated components, with usually close to \SI{95}{\percent} energy contained in all estimated components in the medium environment.
In the large environment, the fixed number of $K=20$ components is not sufficient, capturing usually below \SI{80}{\percent} or the signal energy.
The significantly lower percentage of \SI{82}{\percent} for M4 can be explained by the proximity to metal shelves, from which a larger number of scattered components can be expected, resulting in more diffuse components not well represented by the deterministic \gls{mpc} model.
Similarly, locations L3 and L5 which experience strong obstructions due to the metal shelves only capture below \SI{40}{\percent} of the total signal energy on average, which is also attributable to a larger number of diffuse components not well represented by the model.


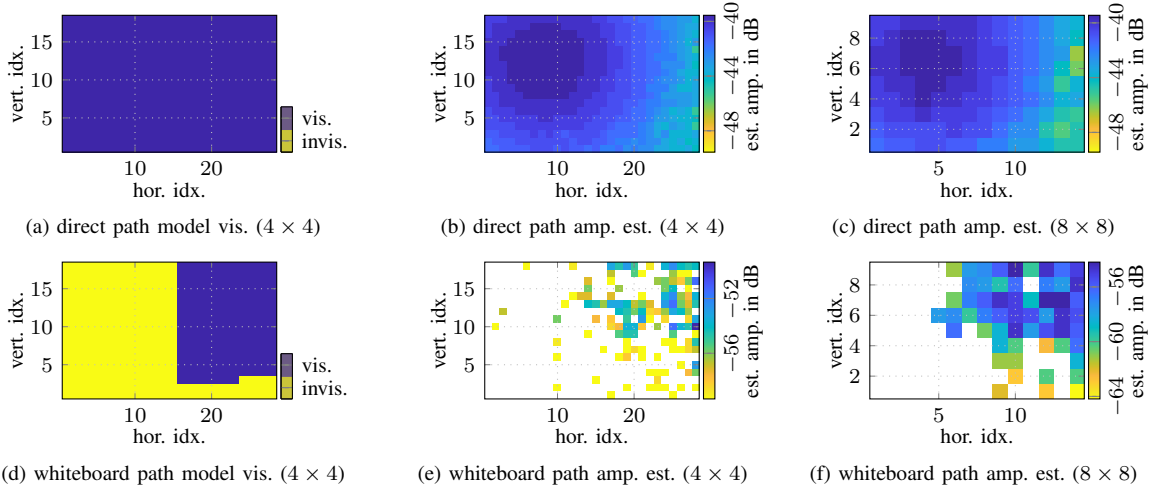
\begin{figure*}[t]
  \vspace{-2mm}
  \centering
  \subfloat[direct path model vis. ($4\times 4$)]{
    \setlength{\figurewidth}{0.28\textwidth}
    \setlength{\figureheight}{0.1\textwidth}
    \def\datapath{./figures/results/model_visibility/portal2_pos1_4x4/VA1_directpath_path__model_vis_img/}

%
%
 
\pgfplotsset{every axis/.append style={
  label style={font=\footnotesize},
  legend style={font=\footnotesize},
  tick label style={font=\footnotesize},
  xticklabel={
    \ifdim \tick pt < 0pt
      \pgfmathparse{abs(\tick)}%
      \llap{$-{}$}\pgfmathprintnumber{\pgfmathresult}
   \else
      \pgfmathprintnumber{\tick}
   \fi
}}}
 
\begin{tikzpicture}

\begin{axis}[%
width=\figurewidth,
height=\figureheight,
at={(0\figurewidth,0\figureheight)},
scale only axis,
point meta min=0,
point meta max=1,
axis on top,
xmin=0.5,
xmax=28.5,
xlabel={hor. idx.},
ymin=0.5,
ymax=18.5,
ylabel={vert. idx.},
axis background/.style={fill=white},
xmajorgrids,
ymajorgrids,
major tick length=0.75mm,
grid style=dotted,
legend style={legend cell align=left, align=left, draw=white!15!black},
 unit vector ratio=1 1 1,
colormap={mymap}{[1pt] rgb(0pt)=(0.9769,0.9839,0.0805); rgb(1pt)=(0.2422,0.1504,0.6603)},
colorbar sampled,
colorbar style={samples=3,at={(1.02,0)},xticklabel pos=right,anchor=south west,major tick length=1mm,height=6mm,width=1.5mm,ytick={0.25,0.75},yticklabels={invis.,vis.},xticklabel style={rotate=0,font=\footnotesize,xshift=0mm},xtick distance={1}},
ylabel style={yshift=-1.5mm},
xlabel style={yshift=1.5mm},
]

\addplot [forget plot] graphics [xmin=0.5, xmax=28.5, ymin=0.5, ymax=18.5] {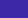};
\end{axis}
\end{tikzpicture}%
    \label{fig:meas:tug:static:medium:pos1:VA1:vis}
  }
  \subfloat[direct path amp. est. ($4\times 4$)]{
    \setlength{\figurewidth}{0.28\textwidth}
    \setlength{\figureheight}{0.1\textwidth}
    \def\datapath{./figures/results/sbl/2022-08_VNA_full_array/array-Nx1_Ny112_Nz75_signal-bw500MHz_fc6950MHz_subarray_sbl-Nx1_Ny4_Nz4/portal2_pos1/VA1_MPC_amplitude_PL-comp_img/}

%
%
 
\pgfplotsset{every axis/.append style={
  label style={font=\footnotesize},
  legend style={font=\footnotesize},
  tick label style={font=\footnotesize},
  xticklabel={
    \ifdim \tick pt < 0pt
      \pgfmathparse{abs(\tick)}%
      \llap{$-{}$}\pgfmathprintnumber{\pgfmathresult}
   \else
      \pgfmathprintnumber{\tick}
   \fi
}}}
 
\begin{tikzpicture}

\begin{axis}[%
width=\figurewidth,
height=\figureheight,
at={(0\figurewidth,0\figureheight)},
scale only axis,
point meta min=-49.5584064092297,
point meta max=-39.5584064092297,
axis on top,
xmin=0.5,
xmax=28.5,
xlabel={hor. idx.},
ymin=0.5,
ymax=18.5,
ylabel={vert. idx.},
axis background/.style={fill=white},
xmajorgrids,
ymajorgrids,
major tick length=0.75mm,
grid style=dotted,
legend style={legend cell align=left, align=left, draw=white!15!black},
unit vector ratio=1 1 1,
colormap={mymap}{[1pt] rgb(0pt)=(0.9769,0.9839,0.0805); rgb(1pt)=(0.960065,0.919778,0.138513); rgb(2pt)=(0.972487,0.851822,0.170039); rgb(3pt)=(0.99583,0.784087,0.206396); rgb(4pt)=(0.974183,0.729983,0.240235); rgb(5pt)=(0.876009,0.73807,0.163478); rgb(6pt)=(0.752635,0.764204,0.175091); rgb(7pt)=(0.607761,0.788896,0.26853); rgb(8pt)=(0.450713,0.802217,0.387143); rgb(9pt)=(0.307896,0.798696,0.505248); rgb(10pt)=(0.206843,0.781061,0.60543); rgb(11pt)=(0.135809,0.757291,0.6878); rgb(12pt)=(0.0091913,0.732587,0.762583); rgb(13pt)=(0.054713,0.700157,0.83063); rgb(14pt)=(0.115917,0.657465,0.881052); rgb(15pt)=(0.146526,0.6075,0.909004); rgb(16pt)=(0.175852,0.553443,0.948983); rgb(17pt)=(0.188526,0.494822,0.9855); rgb(18pt)=(0.249787,0.428622,0.997791); rgb(19pt)=(0.276183,0.365713,0.9825); rgb(20pt)=(0.2814,0.305435,0.945196); rgb(21pt)=(0.276739,0.244439,0.885183); rgb(22pt)=(0.262948,0.193548,0.783748); rgb(23pt)=(0.2422,0.1504,0.6603)},
colorbar,
colorbar style={at={(1.02,0)},xticklabel pos=right,anchor=south west,major tick length=1mm,width=1.5mm,height=\figureheight,yticklabel style={rotate=90,font=\footnotesize,xshift=0mm},ylabel=est. amp. in dB, ylabel style={yshift=2mm}, ytick distance={4}},
ylabel style={yshift=-1.5mm},
xlabel style={yshift=1.5mm},
]
\addplot [forget plot] graphics [xmin=0.5, xmax=28.5, ymin=0.5, ymax=18.5] {\datapath/VA1_MPC_amplitude_PL-comp_img-1.png};
\end{axis}
\end{tikzpicture}%
    \label{fig:meas:tug:static:medium:pos1:VA1:4x4}
  }
  \subfloat[direct path amp. est. ($8\times 8$)]{
    \setlength{\figurewidth}{0.28\textwidth}
    \setlength{\figureheight}{0.1\textwidth}
    \def\datapath{./figures/results/sbl/2022-08_VNA_full_array/array-Nx1_Ny112_Nz75_signal-bw500MHz_fc6950MHz_subarray_sbl-Nx1_Ny8_Nz8/portal2_pos1/VA1_MPC_amplitude_PL-comp_img/}

%
%
 
\pgfplotsset{every axis/.append style={
  label style={font=\footnotesize},
  legend style={font=\footnotesize},
  tick label style={font=\footnotesize},
  xticklabel={
    \ifdim \tick pt < 0pt
      \pgfmathparse{abs(\tick)}%
      \llap{$-{}$}\pgfmathprintnumber{\pgfmathresult}
   \else
      \pgfmathprintnumber{\tick}
   \fi
}}}
 
\begin{tikzpicture}

\begin{axis}[%
width=\figurewidth,
height=\figureheight,
at={(0\figurewidth,0\figureheight)},
scale only axis,
point meta min=-49.5584064092297,
point meta max=-39.5584064092297,
axis on top,
xmin=0.5,
xmax=28.5,
xlabel={hor. idx.},
ymin=0.5,
ymax=18.5,
ylabel={vert. idx.},
axis background/.style={fill=white},
xmajorgrids,
ymajorgrids,
major tick length=0.75mm,
grid style=dotted,
legend style={legend cell align=left, align=left, draw=white!15!black},
unit vector ratio=1 1 1,
colormap={mymap}{[1pt] rgb(0pt)=(0.9769,0.9839,0.0805); rgb(1pt)=(0.960065,0.919778,0.138513); rgb(2pt)=(0.972487,0.851822,0.170039); rgb(3pt)=(0.99583,0.784087,0.206396); rgb(4pt)=(0.974183,0.729983,0.240235); rgb(5pt)=(0.876009,0.73807,0.163478); rgb(6pt)=(0.752635,0.764204,0.175091); rgb(7pt)=(0.607761,0.788896,0.26853); rgb(8pt)=(0.450713,0.802217,0.387143); rgb(9pt)=(0.307896,0.798696,0.505248); rgb(10pt)=(0.206843,0.781061,0.60543); rgb(11pt)=(0.135809,0.757291,0.6878); rgb(12pt)=(0.0091913,0.732587,0.762583); rgb(13pt)=(0.054713,0.700157,0.83063); rgb(14pt)=(0.115917,0.657465,0.881052); rgb(15pt)=(0.146526,0.6075,0.909004); rgb(16pt)=(0.175852,0.553443,0.948983); rgb(17pt)=(0.188526,0.494822,0.9855); rgb(18pt)=(0.249787,0.428622,0.997791); rgb(19pt)=(0.276183,0.365713,0.9825); rgb(20pt)=(0.2814,0.305435,0.945196); rgb(21pt)=(0.276739,0.244439,0.885183); rgb(22pt)=(0.262948,0.193548,0.783748); rgb(23pt)=(0.2422,0.1504,0.6603)},
colorbar,
colorbar style={at={(1.02,0)},xticklabel pos=right,anchor=south west,major tick length=1mm,width=1.5mm,height=\figureheight,yticklabel style={rotate=90,font=\footnotesize,xshift=0mm},ylabel=est. amp. in dB, ylabel style={yshift=2mm}, ytick distance={4}},
ylabel style={yshift=-1.5mm},
xlabel style={yshift=1.5mm},
]
\addplot [forget plot] graphics [xmin=0.5, xmax=28.5, ymin=0.5, ymax=18.5] {\datapath/VA1_MPC_amplitude_PL-comp_img-1.png};
\end{axis}
\end{tikzpicture}%
    \label{fig:meas:tug:static:medium:pos1:VA1:8x8}
  }\vspace{-3mm}
  
  \subfloat[whiteboard path model vis. ($4\times 4$)]{
    \setlength{\figurewidth}{0.28\textwidth}
    \setlength{\figureheight}{0.1\textwidth}
    \def\datapath{./figures/results/model_visibility/portal2_pos1_4x4/VA5_whiteboard_path_98_model_vis_img/}

%
%

\pgfplotsset{every axis/.append style={
label style={font=\footnotesize},
legend style={font=\footnotesize},
tick label style={font=\footnotesize},
xticklabel={
  \ifdim \tick pt < 0pt
  \pgfmathparse{abs(\tick)}%
  \llap{$-{}$}\pgfmathprintnumber{\pgfmathresult}
  \else
  \pgfmathprintnumber{\tick}
  \fi
}}}

\begin{tikzpicture}

\begin{axis}[%
width=\figurewidth,
height=\figureheight,
at={(0\figurewidth,0\figureheight)},
scale only axis,
point meta min=0,
point meta max=1,
axis on top,
xmin=0.5,
xmax=28.5,
xlabel={hor. idx.},
ymin=0.5,
ymax=18.5,
ylabel={vert. idx.},
axis background/.style={fill=white},
xmajorgrids,
ymajorgrids,
major tick length=0.75mm,
grid style=dotted,
legend style={legend cell align=left, align=left, draw=white!15!black},
 unit vector ratio=1 1 1,
colormap={mymap}{[1pt] rgb(0pt)=(0.98,0.98,0.08); rgb(1pt)=(0.24,0.15,0.66)},
colorbar sampled,
colorbar style={samples=3,at={(1.02,0)},xticklabel pos=right,anchor=south west,major tick length=1mm,height=6mm,width=1.5mm,ytick={0.25,0.75},yticklabels={invis.,vis.},xticklabel style={rotate=0,font=\footnotesize,xshift=0mm},xtick distance={1}},
xlabel style={yshift=1.5mm},
ylabel style={yshift=-1.5mm},
]
\addplot [forget plot] graphics [xmin=0.5, xmax=28.5, ymin=0.5, ymax=18.5] {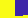};
\end{axis}
\end{tikzpicture}%
    \label{fig:meas:tug:static:medium:pos1:VA5:vis}
  }
  \subfloat[whiteboard path amp. est. ($4\times 4$)]{
    \setlength{\figurewidth}{0.28\textwidth}
    \setlength{\figureheight}{0.1\textwidth}
    \def\datapath{./figures/results/sbl/2022-08_VNA_full_array/array-Nx1_Ny112_Nz75_signal-bw500MHz_fc6950MHz_subarray_sbl-Nx1_Ny4_Nz4/portal2_pos1/VA5_MPC_amplitude_PL-comp_img/}

%
%

\pgfplotsset{every axis/.append style={
label style={font=\footnotesize},
legend style={font=\footnotesize},
tick label style={font=\footnotesize},
xticklabel={
  \ifdim \tick pt < 0pt
  \pgfmathparse{abs(\tick)}%
  \llap{$-{}$}\pgfmathprintnumber{\pgfmathresult}
  \else
  \pgfmathprintnumber{\tick}
  \fi
}}}

\begin{tikzpicture}

\begin{axis}[%
width=\figurewidth,
height=\figureheight,
at={(0\figurewidth,0\figureheight)},
scale only axis,
point meta min=-59.3386506115294,
point meta max=-49.3386506115294,
axis on top,
xmin=0.5,
xmax=28.5,
xlabel={hor. idx.},
ymin=0.5,
ymax=18.5,
ylabel={vert. idx.},
axis background/.style={fill=white},
xmajorgrids,
ymajorgrids,
major tick length=0.75mm,
grid style=dotted,
legend style={legend cell align=left, align=left, draw=white!15!black},
 unit vector ratio=1 1 1,
colormap={mymap}{[1pt] rgb(0pt)=(0.9769,0.9839,0.0805); rgb(1pt)=(0.960065,0.919778,0.138513); rgb(2pt)=(0.972487,0.851822,0.170039); rgb(3pt)=(0.99583,0.784087,0.206396); rgb(4pt)=(0.974183,0.729983,0.240235); rgb(5pt)=(0.876009,0.73807,0.163478); rgb(6pt)=(0.752635,0.764204,0.175091); rgb(7pt)=(0.607761,0.788896,0.26853); rgb(8pt)=(0.450713,0.802217,0.387143); rgb(9pt)=(0.307896,0.798696,0.505248); rgb(10pt)=(0.206843,0.781061,0.60543); rgb(11pt)=(0.135809,0.757291,0.6878); rgb(12pt)=(0.0091913,0.732587,0.762583); rgb(13pt)=(0.054713,0.700157,0.83063); rgb(14pt)=(0.115917,0.657465,0.881052); rgb(15pt)=(0.146526,0.6075,0.909004); rgb(16pt)=(0.175852,0.553443,0.948983); rgb(17pt)=(0.188526,0.494822,0.9855); rgb(18pt)=(0.249787,0.428622,0.997791); rgb(19pt)=(0.276183,0.365713,0.9825); rgb(20pt)=(0.2814,0.305435,0.945196); rgb(21pt)=(0.276739,0.244439,0.885183); rgb(22pt)=(0.262948,0.193548,0.783748); rgb(23pt)=(0.2422,0.1504,0.6603)},
colorbar,
colorbar style={at={(1.02,0)},xticklabel pos=right,anchor=south west,major tick length=1mm,width=1.5mm,height=\figureheight,yticklabel style={rotate=90,font=\footnotesize,xshift=0mm},ylabel=est. amp. in dB, ylabel style={yshift=2mm}, ytick distance={4}},
ylabel style={yshift=-1.5mm},
xlabel style={yshift=1.5mm},
]
\addplot [forget plot] graphics [xmin=0.5, xmax=28.5, ymin=0.5, ymax=18.5] {\datapath/VA5_MPC_amplitude_PL-comp_img-1.png};
\end{axis}
\end{tikzpicture}%
    \label{fig:meas:tug:static:medium:pos1:VA5:4x4}
  }
  \subfloat[whiteboard path amp. est. ($8\times 8$)]{
    \setlength{\figurewidth}{0.28\textwidth}
    \setlength{\figureheight}{0.1\textwidth}
    \def\datapath{./figures/results/sbl/2022-08_VNA_full_array/array-Nx1_Ny112_Nz75_signal-bw500MHz_fc6950MHz_subarray_sbl-Nx1_Ny8_Nz8/portal2_pos1/VA5_MPC_amplitude_PL-comp_img/}

%
%

\pgfplotsset{every axis/.append style={
label style={font=\footnotesize},
legend style={font=\footnotesize},
tick label style={font=\footnotesize},
xticklabel={
  \ifdim \tick pt < 0pt
  \pgfmathparse{abs(\tick)}%
  \llap{$-{}$}\pgfmathprintnumber{\pgfmathresult}
  \else
  \pgfmathprintnumber{\tick}
  \fi
}}}

\begin{tikzpicture}

\begin{axis}[%
width=\figurewidth,
height=\figureheight,
at={(0\figurewidth,0\figureheight)},
scale only axis,
point meta min=-59.3386506115294,
point meta max=-49.3386506115294,
axis on top,
xmin=0.5,
xmax=28.5,
xlabel={hor. idx.},
ymin=0.5,
ymax=18.5,
ylabel={vert. idx.},
axis background/.style={fill=white},
xmajorgrids,
ymajorgrids,
major tick length=0.75mm,
grid style=dotted,
legend style={legend cell align=left, align=left, draw=white!15!black},
 unit vector ratio=1 1 1,
colormap={mymap}{[1pt] rgb(0pt)=(0.9769,0.9839,0.0805); rgb(1pt)=(0.960065,0.919778,0.138513); rgb(2pt)=(0.972487,0.851822,0.170039); rgb(3pt)=(0.99583,0.784087,0.206396); rgb(4pt)=(0.974183,0.729983,0.240235); rgb(5pt)=(0.876009,0.73807,0.163478); rgb(6pt)=(0.752635,0.764204,0.175091); rgb(7pt)=(0.607761,0.788896,0.26853); rgb(8pt)=(0.450713,0.802217,0.387143); rgb(9pt)=(0.307896,0.798696,0.505248); rgb(10pt)=(0.206843,0.781061,0.60543); rgb(11pt)=(0.135809,0.757291,0.6878); rgb(12pt)=(0.0091913,0.732587,0.762583); rgb(13pt)=(0.054713,0.700157,0.83063); rgb(14pt)=(0.115917,0.657465,0.881052); rgb(15pt)=(0.146526,0.6075,0.909004); rgb(16pt)=(0.175852,0.553443,0.948983); rgb(17pt)=(0.188526,0.494822,0.9855); rgb(18pt)=(0.249787,0.428622,0.997791); rgb(19pt)=(0.276183,0.365713,0.9825); rgb(20pt)=(0.2814,0.305435,0.945196); rgb(21pt)=(0.276739,0.244439,0.885183); rgb(22pt)=(0.262948,0.193548,0.783748); rgb(23pt)=(0.2422,0.1504,0.6603)},
colorbar,
colorbar style={at={(1.02,0)},xticklabel pos=right,anchor=south west,major tick length=1mm,width=1.5mm,height=\figureheight,yticklabel style={rotate=90,font=\footnotesize,xshift=0mm},ylabel=est. amp. in dB, ylabel style={yshift=2mm}, ytick distance={4}},
ylabel style={yshift=-1.5mm},
xlabel style={yshift=1.5mm},
]
\addplot [forget plot] graphics [xmin=0.5, xmax=28.5, ymin=0.5, ymax=18.5] {\datapath/VA5_MPC_amplitude_PL-comp_img-1.png};
\end{axis}
\end{tikzpicture}%
    \label{fig:meas:tug:static:medium:pos1:VA5:8x8}
  }%

  \caption{Estimated amplitudes obtained with the \gls{sbl}-algorithm \cite{HansenSAM2014} for $(4\times 4)$ and $(8\times 8)$ subarrays and 500~MHz bandwidth. Data association is performed for direct path and whiteboard components. 
  }\label{fig:meas:tug:static:medium}%
  \vspace{\belowFigureMargin}
\end{figure*}

\subsubsection{Component visibility}
To analyze the visibility and spatial amplitude distribution of estimated components, we rely on the environment models shown in Fig.~\ref{fig:env:corridor} and \ref{fig:env:corridor} to compute image source positions for expected \glspl{mpc}.
For each subarray, the estimates obtained from the \gls{sbl}-algorithm (component delay, azimuth, and elevation) are associated with the components corresponding to the computed image sources using the \gls{da} algorithm from \cite{MeissnerArxiv2013}.

\paragraph{Medium environment}
The estimated amplitudes per subarray associated with modeled components are shown alongside the computed geometric visibility using the model (and ray tracing) in Fig.~\ref{fig:meas:tug:static:medium}, shown by the example of M1 and selected components. 
The direct path is visible at all subarrays independent of their size, showing a similar distribution of estimated amplitudes (see Fig.~\ref{fig:meas:tug:static:medium:pos1:VA1:4x4} and \ref{fig:meas:tug:static:medium:pos1:VA1:8x8}).
The combined effect of the antenna gain patterns is again observable, with the shown estimated amplitudes compensated for the distance-dependent path loss.  
The whiteboard component is only visible in a rectangular area in the upper right corner of the full array (see Figure~\ref{fig:meas:tug:static:medium:pos1:VA5:vis}), which represents the subarray regions where components were found that could be associated to the corresponding image source.
The differences between small and large subarray size in Fig.~\ref{fig:meas:tug:static:medium:pos1:VA5:4x4} and \ref{fig:meas:tug:static:medium:pos1:VA5:8x8} can be attributed to the increased array gain of the larger ones, as components with lower \gls{snr} can be detected.


\begin{figure*}
  \centering  
    \subfloat[direct path model vis. ($4\times 4$)]{\hspace{-5mm}
      \setlength{\figurewidth}{0.2\textwidth}
      \setlength{\figureheight}{0.11\textwidth}
      \def\datapath{./figures/results/sbl/2022-11_VNA_corridor/portal2_pos1/VA1_directpath_path__model_vis_img/}

%
%
 
\pgfplotsset{every axis/.append style={
  label style={font=\footnotesize},
  legend style={font=\footnotesize},
  tick label style={font=\footnotesize},
  xticklabel={
    \ifdim \tick pt < 0pt
      \pgfmathparse{abs(\tick)}%
      \llap{$-{}$}\pgfmathprintnumber{\pgfmathresult}
   \else
      \pgfmathprintnumber{\tick}
   \fi
}}}
 
\begin{tikzpicture}

\begin{axis}[%
width=\figurewidth,
height=\figureheight,
at={(0\figurewidth,0\figureheight)},
scale only axis,
point meta min=0,
point meta max=1,
axis on top,
xmin=0.5,
xmax=28.5,
xlabel={hor. idx.},
ymin=0.5,
ymax=18.5,
ylabel={vert. idx.},
axis background/.style={fill=white},
xmajorgrids,
ymajorgrids,
major tick length=0.75mm,
grid style=dotted,
legend style={legend cell align=left, align=left, draw=white!15!black},
 unit vector ratio=1 1 1,
colormap={mymap}{[1pt] rgb(0pt)=(0.9769,0.9839,0.0805); rgb(1pt)=(0.2422,0.1504,0.6603)},
colorbar sampled,
colorbar style={samples=3,at={(1.02,0)},xticklabel pos=right,anchor=south west,major tick length=1mm,height=6mm,width=1.5mm,ytick={0.25,0.75},yticklabels={invis.,vis.},xticklabel style={rotate=0,font=\footnotesize,xshift=0mm},xtick distance={1}},
ylabel style={yshift=-1.5mm},
xlabel style={yshift=1.5mm},
]

\addplot [forget plot] graphics [xmin=0.5, xmax=28.5, ymin=0.5, ymax=18.5] {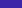};
\end{axis}
\end{tikzpicture}%
      \label{fig:meas:tug:static:large:pos1:VA1:vis}
    }
    \subfloat[right wall path model vis.]{\hspace{-5mm}
      \setlength{\figurewidth}{0.2\textwidth}
      \setlength{\figureheight}{0.11\textwidth}
      \def\datapath{./figures/results/sbl/2022-11_VNA_corridor/portal2_pos1/VA5_wallnorth_path_93_model_vis_img/}

%
%
 
\pgfplotsset{every axis/.append style={
  label style={font=\footnotesize},
  legend style={font=\footnotesize},
  tick label style={font=\footnotesize},
  xticklabel={
    \ifdim \tick pt < 0pt
      \pgfmathparse{abs(\tick)}%
      \llap{$-{}$}\pgfmathprintnumber{\pgfmathresult}
   \else
      \pgfmathprintnumber{\tick}
   \fi
}}}
 
\begin{tikzpicture}

\begin{axis}[%
width=\figurewidth,
height=\figureheight,
at={(0\figurewidth,0\figureheight)},
scale only axis,
point meta min=0,
point meta max=1,
axis on top,
xmin=0.5,
xmax=22.5,
xlabel={hor. idx.},
ymin=0.5,
ymax=8.5,
ylabel={vert. idx.},
axis background/.style={fill=white},
xmajorgrids,
ymajorgrids,
major tick length=0.75mm,
grid style=dotted,
legend style={legend cell align=left, align=left, draw=white!15!black},
 unit vector ratio=1 1 1,
colormap={mymap}{[1pt] rgb(0pt)=(0.9769,0.9839,0.0805); rgb(1pt)=(0.2422,0.1504,0.6603)},
colorbar sampled,
colorbar style={samples=3,at={(1.02,0)},xticklabel pos=right,anchor=south west,major tick length=1mm,height=6mm,width=1.5mm,ytick={0.25,0.75},yticklabels={invis.,vis.},xticklabel style={rotate=0,font=\footnotesize,xshift=0mm},xtick distance={1}},
ylabel style={yshift=-1.5mm},
xlabel style={yshift=1.5mm},
]
\addplot [forget plot] graphics [xmin=0.5, xmax=22.5, ymin=0.5, ymax=8.5] {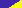};
\end{axis}
\end{tikzpicture}%
      \label{fig:meas:tug:static:large:pos1:VA5:vis}
    }
    \subfloat[left wall path model vis. ($4\times 4$)]{\hspace{-5mm}
      \setlength{\figurewidth}{0.2\textwidth}
      \setlength{\figureheight}{0.11\textwidth}
      \def\datapath{./figures/results/sbl/2022-11_VNA_corridor/portal2_pos1/multiple_segments_15_16_17_92_model_vis_img/}

%
%
 
\pgfplotsset{every axis/.append style={
  label style={font=\footnotesize},
  legend style={font=\footnotesize},
  tick label style={font=\footnotesize},
  xticklabel={
    \ifdim \tick pt < 0pt
      \pgfmathparse{abs(\tick)}%
      \llap{$-{}$}\pgfmathprintnumber{\pgfmathresult}
   \else
      \pgfmathprintnumber{\tick}
   \fi
}}}
 
\begin{tikzpicture}

\begin{axis}[%
width=\figurewidth,
height=\figureheight,
at={(0\figurewidth,0\figureheight)},
scale only axis,
point meta min=0,
point meta max=1,
axis on top,
xmin=0.5,
xmax=22.5,
xlabel={hor. idx.},
ymin=0.5,
ymax=8.5,
ylabel={vert. idx.},
axis background/.style={fill=white},
xmajorgrids,
ymajorgrids,
major tick length=0.75mm,
grid style=dotted,
legend style={legend cell align=left, align=left, draw=white!15!black},
 unit vector ratio=1 1 1,
colormap={mymap}{[1pt] rgb(0pt)=(0.9769,0.9839,0.0805); rgb(1pt)=(0.2422,0.1504,0.6603)},
colorbar sampled,
colorbar style={samples=3,at={(1.02,0)},xticklabel pos=right,anchor=south west,major tick length=1mm,height=6mm,width=1.5mm,ytick={0.25,0.75},yticklabels={invis.,vis.},xticklabel style={rotate=0,font=\footnotesize,xshift=0mm},xtick distance={1}},
ylabel style={yshift=-1.5mm},
xlabel style={yshift=1.5mm},
]
\addplot [forget plot] graphics [xmin=0.5, xmax=22.5, ymin=0.5, ymax=8.5] {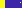};
\end{axis}
\end{tikzpicture}%
      \label{fig:meas:tug:static:large:pos1:VA4:vis}
    }
  
  \subfloat[direct path amp. est. ($4\times 4$)]{\hspace{-5mm}
    \setlength{\figurewidth}{0.2\textwidth}
    \setlength{\figureheight}{0.11\textwidth}
    \def\datapath{./figures/results/sbl/2022-11_VNA_corridor/portal2_pos1/VA1_MPC_amplitude_PL-comp_img/}

%
%
 
\pgfplotsset{every axis/.append style={
  label style={font=\footnotesize},
  legend style={font=\footnotesize},
  tick label style={font=\footnotesize},
  xticklabel={
    \ifdim \tick pt < 0pt
      \pgfmathparse{abs(\tick)}%
      \llap{$-{}$}\pgfmathprintnumber{\pgfmathresult}
   \else
      \pgfmathprintnumber{\tick}
   \fi
}}}
 
\begin{tikzpicture}

\begin{axis}[%
width=\figurewidth,
height=\figureheight,
at={(0\figurewidth,0\figureheight)},
scale only axis,
point meta min=-49.5584064092297,
point meta max=-39.5584064092297,
axis on top,
xmin=0.5,
xmax=28.5,
xlabel={hor. idx.},
ymin=0.5,
ymax=18.5,
ylabel={vert. idx.},
axis background/.style={fill=white},
xmajorgrids,
ymajorgrids,
major tick length=0.75mm,
grid style=dotted,
legend style={legend cell align=left, align=left, draw=white!15!black},
unit vector ratio=1 1 1,
colormap={mymap}{[1pt] rgb(0pt)=(0.9769,0.9839,0.0805); rgb(1pt)=(0.960065,0.919778,0.138513); rgb(2pt)=(0.972487,0.851822,0.170039); rgb(3pt)=(0.99583,0.784087,0.206396); rgb(4pt)=(0.974183,0.729983,0.240235); rgb(5pt)=(0.876009,0.73807,0.163478); rgb(6pt)=(0.752635,0.764204,0.175091); rgb(7pt)=(0.607761,0.788896,0.26853); rgb(8pt)=(0.450713,0.802217,0.387143); rgb(9pt)=(0.307896,0.798696,0.505248); rgb(10pt)=(0.206843,0.781061,0.60543); rgb(11pt)=(0.135809,0.757291,0.6878); rgb(12pt)=(0.0091913,0.732587,0.762583); rgb(13pt)=(0.054713,0.700157,0.83063); rgb(14pt)=(0.115917,0.657465,0.881052); rgb(15pt)=(0.146526,0.6075,0.909004); rgb(16pt)=(0.175852,0.553443,0.948983); rgb(17pt)=(0.188526,0.494822,0.9855); rgb(18pt)=(0.249787,0.428622,0.997791); rgb(19pt)=(0.276183,0.365713,0.9825); rgb(20pt)=(0.2814,0.305435,0.945196); rgb(21pt)=(0.276739,0.244439,0.885183); rgb(22pt)=(0.262948,0.193548,0.783748); rgb(23pt)=(0.2422,0.1504,0.6603)},
colorbar,
colorbar style={at={(1.02,0)},xticklabel pos=right,anchor=south west,major tick length=1mm,width=1.5mm,height=\figureheight,yticklabel style={rotate=90,font=\footnotesize,xshift=0mm},ylabel=est. amp. in dB, ylabel style={yshift=2mm}, ytick distance={4}},
ylabel style={yshift=-1.5mm},
xlabel style={yshift=1.5mm},
]
\addplot [forget plot] graphics [xmin=0.5, xmax=28.5, ymin=0.5, ymax=18.5] {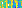};
\end{axis}
\end{tikzpicture}%
    \label{fig:meas:tug:static:large:pos1:VA1:4x4}
  }
  \subfloat[right wall path amp. est. ($4\times 4$)]{\hspace{-5mm}
    \setlength{\figurewidth}{0.2\textwidth}
    \setlength{\figureheight}{0.11\textwidth}
    \def\datapath{./figures/results/sbl/2022-11_VNA_corridor/portal2_pos1/VA5_MPC_amplitude_PL-comp_img/}

%
%

\pgfplotsset{every axis/.append style={
label style={font=\footnotesize},
legend style={font=\footnotesize},
tick label style={font=\footnotesize},
xticklabel={
  \ifdim \tick pt < 0pt
  \pgfmathparse{abs(\tick)}%
  \llap{$-{}$}\pgfmathprintnumber{\pgfmathresult}
  \else
  \pgfmathprintnumber{\tick}
  \fi
}}}

\begin{tikzpicture}

\begin{axis}[%
width=\figurewidth,
height=\figureheight,
at={(0\figurewidth,0\figureheight)},
scale only axis,
point meta min=-59.3386506115294,
point meta max=-49.3386506115294,
axis on top,
xmin=0.5,
xmax=28.5,
xlabel={hor. idx.},
ymin=0.5,
ymax=18.5,
ylabel={vert. idx.},
axis background/.style={fill=white},
xmajorgrids,
ymajorgrids,
major tick length=0.75mm,
grid style=dotted,
legend style={legend cell align=left, align=left, draw=white!15!black},
 unit vector ratio=1 1 1,
colormap={mymap}{[1pt] rgb(0pt)=(0.9769,0.9839,0.0805); rgb(1pt)=(0.960065,0.919778,0.138513); rgb(2pt)=(0.972487,0.851822,0.170039); rgb(3pt)=(0.99583,0.784087,0.206396); rgb(4pt)=(0.974183,0.729983,0.240235); rgb(5pt)=(0.876009,0.73807,0.163478); rgb(6pt)=(0.752635,0.764204,0.175091); rgb(7pt)=(0.607761,0.788896,0.26853); rgb(8pt)=(0.450713,0.802217,0.387143); rgb(9pt)=(0.307896,0.798696,0.505248); rgb(10pt)=(0.206843,0.781061,0.60543); rgb(11pt)=(0.135809,0.757291,0.6878); rgb(12pt)=(0.0091913,0.732587,0.762583); rgb(13pt)=(0.054713,0.700157,0.83063); rgb(14pt)=(0.115917,0.657465,0.881052); rgb(15pt)=(0.146526,0.6075,0.909004); rgb(16pt)=(0.175852,0.553443,0.948983); rgb(17pt)=(0.188526,0.494822,0.9855); rgb(18pt)=(0.249787,0.428622,0.997791); rgb(19pt)=(0.276183,0.365713,0.9825); rgb(20pt)=(0.2814,0.305435,0.945196); rgb(21pt)=(0.276739,0.244439,0.885183); rgb(22pt)=(0.262948,0.193548,0.783748); rgb(23pt)=(0.2422,0.1504,0.6603)},
colorbar,
colorbar style={at={(1.02,0)},xticklabel pos=right,anchor=south west,major tick length=1mm,width=1.5mm,height=\figureheight,yticklabel style={rotate=90,font=\footnotesize,xshift=0mm},ylabel=est. amp. in dB, ylabel style={yshift=2mm}, ytick distance={4}},
ylabel style={yshift=-1.5mm},
xlabel style={yshift=1.5mm},
]
\addplot [forget plot] graphics [xmin=0.5, xmax=28.5, ymin=0.5, ymax=18.5] {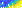};
\end{axis}
\end{tikzpicture}%
    \label{fig:meas:tug:static:large:pos1:VA5:4x4}
  }
  \subfloat[left wall path amp. est. ($4\times 4$)]{\hspace{-5mm}
    \setlength{\figurewidth}{0.2\textwidth}
    \setlength{\figureheight}{0.11\textwidth}
    \def\datapath{./figures/results/sbl/2022-11_VNA_corridor/portal2_pos1/VA4_MPC_amplitude_PL-comp_img/}

%
%
 
\pgfplotsset{every axis/.append style={
  label style={font=\footnotesize},
  legend style={font=\footnotesize},
  tick label style={font=\footnotesize},
  xticklabel={
    \ifdim \tick pt < 0pt
      \pgfmathparse{abs(\tick)}%
      \llap{$-{}$}\pgfmathprintnumber{\pgfmathresult}
   \else
      \pgfmathprintnumber{\tick}
   \fi
}}}
 
\begin{tikzpicture}

\begin{axis}[%
width=0.926\figurewidth,
height=\figureheight,
at={(0\figurewidth,0\figureheight)},
scale only axis,
point meta min=-57.7841471666794,
point meta max=-47.7841471666794,
axis on top,
xmin=0.5,
xmax=28.5,
xlabel={hor. idx.},
ymin=0.5,
ymax=18.5,
ylabel={vert. idx.},
axis background/.style={fill=white},
xmajorgrids,
ymajorgrids,
grid style=dotted,
legend style={legend cell align=left, align=left, draw=white!15!black},
 unit vector ratio=1 1 1,
colormap={mymap}{[1pt] rgb(0pt)=(0.9769,0.9839,0.0805); rgb(1pt)=(0.960065,0.919778,0.138513); rgb(2pt)=(0.972487,0.851822,0.170039); rgb(3pt)=(0.99583,0.784087,0.206396); rgb(4pt)=(0.974183,0.729983,0.240235); rgb(5pt)=(0.876009,0.73807,0.163478); rgb(6pt)=(0.752635,0.764204,0.175091); rgb(7pt)=(0.607761,0.788896,0.26853); rgb(8pt)=(0.450713,0.802217,0.387143); rgb(9pt)=(0.307896,0.798696,0.505248); rgb(10pt)=(0.206843,0.781061,0.60543); rgb(11pt)=(0.135809,0.757291,0.6878); rgb(12pt)=(0.0091913,0.732587,0.762583); rgb(13pt)=(0.054713,0.700157,0.83063); rgb(14pt)=(0.115917,0.657465,0.881052); rgb(15pt)=(0.146526,0.6075,0.909004); rgb(16pt)=(0.175852,0.553443,0.948983); rgb(17pt)=(0.188526,0.494822,0.9855); rgb(18pt)=(0.249787,0.428622,0.997791); rgb(19pt)=(0.276183,0.365713,0.9825); rgb(20pt)=(0.2814,0.305435,0.945196); rgb(21pt)=(0.276739,0.244439,0.885183); rgb(22pt)=(0.262948,0.193548,0.783748); rgb(23pt)=(0.2422,0.1504,0.6603)},
colorbar,
colorbar style={at={(1.02,0)},xticklabel pos=right,anchor=south west,major tick length=1mm,width=1.5mm,height=\figureheight,yticklabel style={rotate=0,font=\footnotesize,xshift=0mm},ylabel=est. amp. in dB},
ylabel style={yshift=-1.5mm},
xlabel style={yshift=1.5mm},
]
\addplot [forget plot] graphics [xmin=0.5, xmax=28.5, ymin=0.5, ymax=18.5] {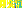};
\end{axis}
\end{tikzpicture}%
    \label{fig:meas:tug:static:large:pos1:VA4:4x4}
  }
  
  \caption{Estimated amplitudes obtained with the \gls{sbl}-algorithm from \cite{HansenSAM2014} for the large environment at position L1. Subarrays of size $(4\times 4)$ are used with $500~\mathrm{MHz}$ bandwidth, comparing two first order reflections.
  }\label{fig:meas:tug:static:large}%
  \vspace{\belowFigureMargin}
\end{figure*}

\paragraph{Large environment} 
Similar results are obtained for the large-sized environment.
At the example of measurement position L1, Fig.~\ref{fig:meas:tug:static:large} shows the results for subarrays of dimension $(4\times 4)$ and a bandwidth of \SI{500}{\mega\hertz}.
Note that the measurement area is smaller due to the position in the environment.
Additionally, the lower carrier frequency of $f_\text{c}=\SI{6}{\giga\hertz}$ results in a slightly larger inter-element spacing.
The figure shows the computed visibility (top row of subplots) and the amplitude estimates obtained with the \gls{sbl} algorithm (bottom row) for the direct path, and the walls to the left and right of the array (as seen from the array).

The direct path (Fig.~\ref{fig:meas:tug:static:large:pos1:VA1:vis} and~\ref{fig:meas:tug:static:large:pos1:VA1:4x4}) is generally in \gls{los} condition for the full \gls{pla} area. 
Nonetheless, variations in the estimated amplitude can be observed, which are likely caused by overlap between the left wall component (see the model in Fig.~\ref{fig:env:corridor} and photo in \ref{fig:env:to2}), as the $(4\times 4)$ \glspl{ura} have a comparably low spatial resolution.
For the reflection via the right wall (Figure~\ref{fig:meas:tug:static:large:pos1:VA5:vis} and~\ref{fig:meas:tug:static:large:pos1:VA5:4x4}), the visibility computed from the environment model shows a triangular region of component invisibility in the bottom right corner of the \gls{pla}, which coincides with the region of low estimated amplitudes of similar shape. 
Note that the geometric shape of the visibility region is due to the measurement area located underneath the right wall (see Fig.~\ref{fig:env:eg}), with the lower wall edge representing the boundary between the visibility regions for the corresponding \gls{mpc}.
The left wall, in turn, shows stronger amplitude fading and less clear correspondence with the computed geometric visibility. 
Again, these variations could be attributed to the geometric configuration of the measurement setup: as the measurement location is close to the corresponding image source, path overlap in the delay and the angle domain again likely causes fading.

\section{Conclusion and Future Work}\label{sec:conclusion}

The synthetic \gls{pla} measurements and the analysis presented in this paper have shown the feasibility and necessity of employing visibility regions for \gls{mpc}-based models in the context of \glspl{pla}. 
Based on the channel measurements, we have shown that it is possible to perform super-resolution channel estimation for subarrays for which local stationarity can be assumed, inherently accounting for visibility. 
This is important in location-aware applications, where the measured channel is used to estimate the user location or infer environment parameters.
Consequently, robust algorithms must consider the limited component visibility and varying number of components and received signal power. 

Future work will deal with algorithms that exploit the described propagation conditions as well as refinement of the outlined channel model that considers visibility on a per-array, or at least per-subarray level.
Furthermore, algorithms for the optimal fusion of local measurements/estimates obtained by subarrays to attain the performance of a corresponding fully coherent \gls{pla} aperture are under development based on the channel measurements.

\section*{Acknowledgment}
The project has received funding from the European Union’s Horizon 2020 research and
innovation programme under grant agreement No 101013425 (Project ``REINDEER").

\balance

\bibliographystyle{IEEEtran}
\bibliography{IEEEabrv,EUCNC}

\begin{thebibliography}{10}
\providecommand{\url}[1]{#1}
\csname url@samestyle\endcsname
\providecommand{\newblock}{\relax}
\providecommand{\bibinfo}[2]{#2}
\providecommand{\BIBentrySTDinterwordspacing}{\spaceskip=0pt\relax}
\providecommand{\BIBentryALTinterwordstretchfactor}{4}
\providecommand{\BIBentryALTinterwordspacing}{\spaceskip=\fontdimen2\font plus
\BIBentryALTinterwordstretchfactor\fontdimen3\font minus
  \fontdimen4\font\relax}
\providecommand{\BIBforeignlanguage}[2]{{%
\expandafter\ifx\csname l@#1\endcsname\relax
\typeout{** WARNING: IEEEtran.bst: No hyphenation pattern has been}%
\typeout{** loaded for the language `#1'. Using the pattern for}%
\typeout{** the default language instead.}%
\else
\language=\csname l@#1\endcsname
\fi
#2}}
\providecommand{\BIBdecl}{\relax}
\BIBdecl

\bibitem{Guerra_Access2022}
D.~W.~M. Guerra and T.~Abr{\~{a}}o, ``Clustered double-scattering channel
  modeling for {XL}-{MIMO} with uniform arrays,'' \emph{{IEEE} Access},
  vol.~10, pp. 20\,173--20\,186, Feb. 2022.

\bibitem{ShaikTC2021}
Z.~H. Shaik, E.~Bj{\"o}rnson, and E.~G. Larsson, ``Mmse-optimal sequential
  processing for cell-free massive mimo with radio stripes,'' \emph{{IEEE}
  Trans. Commun.}, vol.~69, no.~11, pp. 7775--7789, 2021.

\bibitem{DardariJSAC2020}
D.~Dardari, ``Communicating with large intelligent surfaces: Fundamental limits
  and models,'' \emph{{IEEE} J. Sel. Areas Commun.}, vol.~38, no.~11, pp.
  2526--2537, 2020.

\bibitem{BjoernsonSPM2022}
E.~Bj{\"o}rnson, H.~Wymeersch, B.~Matthiesen, P.~Popovski, L.~Sanguinetti, and
  E.~de~Carvalho, ``Reconfigurable intelligent surfaces: A signal processing
  perspective with wireless applications,'' \emph{{IEEE} Signal Process. Mag.},
  vol.~39, no.~2, pp. 135--158, 2022.

\bibitem{VanderPerreAsilomar2019}
L.~{Van der Perre}, E.~G. Larsson, F.~Tufvesson, L.~{De Strycker},
  E.~Bj{\"{o}}rnson, and O.~Edfors, ``{RadioWeaves for efficient connectivity:
  analysis and impact of constraints in actual deployments},'' in \emph{2019
  53rd Asilomar Conference on Signals, Systems, and Computers}.\hskip 1em plus
  0.5em minus 0.4em\relax IEEE, nov 2019, pp. 15--22.

\bibitem{D1_1}
J.~F. Esteban and M.~Truskaller, ``{Use case-driven specifications and
  technical requirements and initial channel model},'' {REINDEER project},
  Deliverable {ICT-52-2020 / D1.1}, Sep. 2021.

\bibitem{GaoAsilomar2013}
X.~Gao, F.~Tufvesson, and O.~Edfors, ``Massive mimo channels - measurements and
  models,'' in \emph{2013 Asilomar Conference on Signals, Systems and
  Computers}, 2013, pp. 280--284.

\bibitem{CarvalhoWC2020}
E.~D. Carvalho, A.~Ali, A.~Amiri, M.~Angjelichinoski, and R.~W. Heath,
  ``Non-stationarities in extra-large-scale massive mimo,'' \emph{IEEE Wireless
  Communications}, vol.~27, no.~4, pp. 74--80, 2020.

\bibitem{FlordelisTWC2020}
J.~Flordelis, X.~Li, O.~Edfors, and F.~Tufvesson, ``Massive {MIMO} extensions
  to the {COST} 2100 channel model: Modeling and validation,'' \emph{{IEEE}
  Trans. Wireless Commun.}, vol.~19, no.~1, pp. 380--394, Oct. 2020.

\bibitem{ChenArxiv2022}
\BIBentryALTinterwordspacing
H.~Chen, A.~Elzanaty, R.~Ghazalian, M.~F. Keskin, R.~J{\"a}ntti, and
  H.~Wymeersch, ``Channel model mismatch analysis for xl-{MIMO} systems from a
  localization perspective,'' 2022. [Online]. Available:
  \url{https://arxiv.org/abs/2205.15417}
\BIBentrySTDinterwordspacing

\bibitem{PedersenTAP2018}
T.~Pedersen, ``Modeling of path arrival rate for in-room radio channels with
  directive antennas,'' \emph{{IEEE} Trans. Antennas Propag.}, vol.~66, no.~9,
  pp. 4791--4805, 2018.

\bibitem{SchubertTAP2013}
F.~M. Schubert, M.~L. Jakobsen, and B.~H. Fleury, ``Non-stationary propagation
  model for scattering volumes with an application to the rural {LMS}
  channel,'' \emph{{IEEE} Trans. Antennas Propag.}, vol.~61, no.~5, pp.
  2817--2828, 2013.

\bibitem{RichterPhD2005}
A.~Richter, ``{Estimation of Radio Channel Parameters: Models and
  Algorithms},'' Ph.D. dissertation, Ilmenau University of Technology, 2005.

\bibitem{SalmiTSP2009}
J.~Salmi, A.~Richter, and V.~Koivunen, ``Detection and tracking of {MIMO}
  propagation path parameters using state-space approach,'' \emph{{IEEE} Trans.
  Signal Process.}, vol.~57, no.~4, pp. 1538--1550, Apr. 2009.

\bibitem{LeitingerJSAC2015}
E.~Leitinger, P.~Meissner, C.~Rudisser, G.~Dumphart, and K.~Witrisal,
  ``Evaluation of position-related information in multipath components for
  indoor positioning,'' \emph{{IEEE} J. Sel. Areas Commun.}, vol.~33, no.~11,
  pp. 2313--2328, Nov. 2015.

\bibitem{johnson1992}
D.~H. Johnson and D.~E. Dudgeon, \emph{Array signal processing: concepts and
  techniques}.\hskip 1em plus 0.5em minus 0.4em\relax Simon \& Schuster, Inc.,
  1992.

\bibitem{KrallPhD2008}
C.~Krall, ``Signal processing for ultra wideband transceivers,'' Ph.D.
  dissertation, Graz University of Technology, 2008.

\bibitem{CostaTAP2009}
J.~R. Costa, C.~R. Medeiros, and C.~A. Fernandes, ``Performance of a crossed
  exponentially tapered slot antenna for uwb systems,'' \emph{{IEEE} Trans.
  Antennas Propag.}, vol.~57, no.~5, pp. 1345--1352, 2009.

\bibitem{HansenSAM2014}
T.~L. Hansen, M.~A. Badiu, B.~H. Fleury, and B.~D. Rao, ``A sparse {B}ayesian
  learning algorithm with dictionary parameter estimation,'' in \emph{2014 IEEE
  8th Sensor Array and Multichannel Signal Processing Workshop (SAM)}.\hskip
  1em plus 0.5em minus 0.4em\relax IEEE, 2014, pp. 385--388.

\bibitem{MeissnerArxiv2013}
P.~Meissner, E.~Leitinger, M.~Fr{\"o}hle, and K.~Witrisal, ``Accurate and
  robust indoor localization systems using ultra-wideband signals,''
  \emph{arXiv preprint arXiv:1304.7928}, 2013.

\end{thebibliography}

\end{document}